\documentclass[a4paper]{article}
\usepackage[english]{babel}
\usepackage[utf8]{inputenc}
\usepackage{graphicx, authblk, natbib, multirow, color, hyphenat, framed, alltt, url, enumitem, amssymb}
\usepackage{mathtools}
\usepackage[nolists]{endfloat}

\def\bSig\mathbf{\Sigma}

\newcommand{\bfalpha}{\mbox{{\boldmath $\alpha$}}}
\newcommand{\bfbeta}{\mbox{{\boldmath $\beta$}}}
\newcommand{\bfgamma}{\mbox{{\boldmath $\gamma$}}}
\newcommand{\bftheta}{\mbox{{\boldmath $\theta$}}}

\newcommand{\bfsigma}{\mbox{{\boldmath $\sigma$}}}
\newcommand{\bfOmega}{\mbox{{\boldmath $\Omega$}}}

\newcommand{\bw}{\mbox{{\boldmath $w$}}}
\newcommand{\bv}{\mbox{{\boldmath $v$}}}
\newcommand{\by}{\mbox{{\boldmath $y$}}}
\newcommand{\bx}{\mbox{{\boldmath $x$}}}

\newcommand{\bbI}{\mbox{\textbf I}}
\newcommand{\bz}{\mbox{{\boldmath $z$}}}

\newcommand{\bK}{\mbox{{\boldmath $K$}}}
\newcommand{\bb}{\mbox{{\boldmath $b$}}}

\newcommand{\bD}{\mbox{{\boldmath $D$}}}
\def\argmax{\mathop{\rm arg\,max}}

\newcommand{\R}{\textsf{R}}
\newcommand{\Lim}[1]{\raisebox{0.5ex}{\scalebox{0.8}{$\displaystyle \lim_{#1}\;$}}}

\newcommand{\hlcom}[1]{\textcolor[rgb]{0.678,0.584,0.686}{\textit{#1}}}%
\newcommand{\hlkwd}[1]{\textcolor[rgb]{0.737,0.353,0.396}{\textbf{#1}}}%

\definecolor{fgcolor}{rgb}{0.345, 0.345, 0.345}
\definecolor{shadecolor}{rgb}{.97, .97, .97}
\definecolor{messagecolor}{rgb}{0, 0, 0}
\definecolor{warningcolor}{rgb}{1, 0, 1}
\definecolor{errorcolor}{rgb}{1, 0, 0}
\newenvironment{knitrout}{}{} 

\makeatletter
\newenvironment{kframe}{%
	\def\at@end@of@kframe{}%
	\ifinner\ifhmode%
	\def\at@end@of@kframe{\end{minipage}}%
\begin{minipage}{\columnwidth}%
	\fi\fi%
	\def\FrameCommand##1{\hskip\@totalleftmargin \hskip-\fboxsep
		\colorbox{shadecolor}{##1}\hskip-\fboxsep
		\hskip-\linewidth \hskip-\@totalleftmargin \hskip\columnwidth}%
	\MakeFramed {\advance\hsize-\width
		\@totalleftmargin\z@ \linewidth\hsize
		\@setminipage}}%
{\par\unskip\endMakeFramed%
	\at@end@of@kframe}
\makeatother

\title{Joint Models with Multiple Longitudinal Outcomes and a Time-to-Event Outcome: a Corrected Two-Stage Approach}

\author[1]{\bfseries Katya Mauff}
\author[2, 3]{\textbf{Ewout Steyerberg}}
\author[4]{\textbf{Isabella Kardys}}
\author[4]{\textbf{Eric Boersma}}
\author[1]{\textbf{Dimitris Rizopoulos}}

\affil[1]{\small{Department of Biostatistics, Erasmus Medical Center, Rotterdam, the Netherlands}}
\affil[2]{\small{Center for Medical Decision Making, Department of Public Health, Erasmus Medical Center, Rotterdam, the Netherlands}}
\affil[3] {\small{Department of Biomedical Data Sciences, Leiden University Medical Center, Leiden, the Netherlands}}
\affil[4]{\small{Department of Cardiology, Erasmus University Medical Center, Rotterdam}}

\date{\today}

\begin{document}
\maketitle
\begin{abstract}
Joint models for longitudinal and survival data have gained a lot of attention in recent years, with the development of myriad extensions to the basic model, including those which allow for multivariate longitudinal data, competing risks and recurrent events. Several software packages are now also available for their implementation. Although mathematically straightforward, the inclusion of multiple longitudinal outcomes in the joint model remains computationally difficult due to the large number of random effects required, which hampers the practical application of this extension. We present a novel approach that enables the fitting of such models with more realistic computational times. The idea behind the approach is to split the estimation of the joint model in two steps; estimating a multivariate mixed model for the longitudinal outcomes, and then using the output from this model to fit the survival submodel. So called two-stage approaches have previously been proposed, and shown to be biased. Our approach differs from the standard version, in that we additionally propose the application of a correction factor, adjusting the estimates obtained such that they more closely resemble those we would expect to find with the multivariate joint model. This correction is based on importance sampling ideas. Simulation studies show that this corrected-two-stage approach works satisfactorily, eliminating the bias while maintaining substantial improvement in computational time, even in more difficult settings. 
\end{abstract}


\section{Introduction}
\label{intro}
Joint models for longitudinal and survival data have become a valuable asset in the toolbox of modern data scientists. After the seminal papers of \cite{faucett.thomas:96} and \cite{wulfsohn.tsiatis:97}, several extensions of these models have been proposed in the literature. These include, amongst others, flexible specification of the longitudinal model \citep{brown.et.al:05}, consideration of competing risks \citep{elashoff.et.al:08, andrinopoulou.et.al:14} and multi-state models \citep{ferrer.et.al:16}, and the calculation of dynamic predictions \citep{proust-lima.taylor:09, rizopoulos:11, rizopoulos.et.al:14, andrinopoulou.rizopoulos:16, rizopoulos.et.al:17, andrinopoulou.et.al:18}. A particularly useful and practical extension is that which allows for the inclusion of multiple longitudinal outcomes \citep{RizGhosh2011, ChiIb2006, BrownIbDeG2005, Lin2002}. In medical settings in particular, data collection is likely to be complex: while the standard joint model allows us to determine the association between a survival outcome and a single longitudinal outcome (biomarker), there are more often than not multiple biomarkers relevant to the event of interest. Extending the univariate joint model to accommodate these multiple longitudinal outcomes allows us to incorporate more information, improving prognostication and enabling us to better make sense of the complex underlying nature of the disease dynamics.
A motivating example of this is the Bio-SHiFT cohort study; a prospective observational study conducted in the Netherlands on chronic heart failure (CHF) patients. The primary focus of the study was to determine whether or not disease progression in individual CHF patients can be assessed using longitudinal measurements of several blood biomarkers \citep{bioshift}. Previous work on this data has focused mainly on the association between each individual biomarker and a single composite event, but it is likely that the predictive value of the biomarkers will be more accurately determined when they are assessed in concert. 

Extension to the multivariate case is mathematically straightforward, and may be easily combined with other extensions, allowing for longitudinal outcomes of varying types; left, right and interval censoring; and the inclusion of competing risks, amongst others. There are also now a number of excellent software packages available, which make for easier implementation of the more complex models. There are however technical challenges which hamper the widespread use of these models. As the number of longitudinal outcomes increases, and thus the number of random effects, standard methods become computationally prohibitive: under a Bayesian approach, the number of parameters to sample becomes unreasonably large, and in the case of maximum likelihood, we are required to numerically approximate the integrals over the random effects, which is challenging in high dimensions. The practical solution most commonly used in such settings is that of the two-stage approach, wherein a multivariate mixed model is first used for the longitudinal outcomes, following which, the output of this model is used to fit a survival submodel. Unfortunately, substantial research on this topic indicates that this approach results in biased estimates \citep{tsiatis.davidian:04, rizopoulos:12, ye.et.al:08b}. In this paper, we propose an adaptation of the simple two-stage approach which eliminates the bias and substantially reduces computational time. We propose the use of a correction factor, based on importance sampling theory \citep[Section~7.9]{press.et.al:07}. This correction factor allows us to re-weight each realization of the MCMC sample obtained from the Bayesian estimation of the two-stage approach, such that the resulting estimates more closely approximate those obtained via the full multivariate joint model. The weights are given by the target distribution (the full posterior distribution of the multivariate joint model), divided by the product of the posterior distributions for each of the two stages, evaluated for each iteration of the MCMC sample. The use of this correction factor alone is not enough to eliminate the bias, but, prior to its application, the two-stage approach is itself modified: where before, in the second stage, only the parameters of the survival submodel were updated, we now also update the random effects. These adaptations combined, achieve unbiased estimates in a fraction of the time required to compute the full multivariate model. 

The rest of the paper is organised as follows: Section~\ref{sec:1} introduces the full multivariate joint model, and Section~\ref{sec:2} discusses the estimation of the model under the Bayesian paradigm. Section~\ref{sec:3} introduces the importance-sampling corrected two-stage approach, and presents the results of a simple simulation, and Section~\ref{sec:4} the importance-sampling corrected two-stage approach with updated random effects. Section~\ref{sec:5} presents the results of a more complex simulation, and finally in Section~\ref{sec:6} we look at an analysis of the Bio-SHiFT data. 


\section{Joint Model Specification} 
\label{sec:1}
We start with a general definition of the framework of multivariate joint models for multiple longitudinal outcomes and an event time. 
Let $\mathcal D_n = \{T_i, T_i^U, \delta_i, \by_i; i = 1, \ldots, n\}$ denote a sample from the target population, where $T_i^*$ denotes the true event time for the $i$-th subject, $T_i$ and $T_i^U$ the observed event times. Then $\delta_i \in \{0, 1, 2, 3\}$ denotes the event indicator, with 0 corresponding to right censoring ($T_i^* > T_i$), 1 to a true event ($T_i^* = T_i$), 2 to left censoring ($T_i^* < T_i$), and 3 to interval censoring ($T_i < T_i^* < T_i^U$). Assuming $K$ longitudinal outcomes we let $\by_{ki}$ denote the $n_{ki} \times 1$ longitudinal response vector for the $k$-th outcome ($k = 1, \ldots, K$) and the $i$-th subject, with elements $y_{kij}$ denoting the value of the $k$-th longitudinal outcome for the $i$-th subject, taken at time point $t_{kij}$, $j = 1, \ldots, n_{ki}$.
To accommodate multivariate longitudinal responses of different types in a unified framework, we postulate a generalized linear mixed effects model. In particular, the conditional distribution of $\by_{ki}$ given a vector of random effects $\bb_{ki}$ is assumed to be a member of the exponential family, with linear predictor given by
\small
\begin{gather}
g_k \bigl [ E \{ y_{ki}(t) \mid \bb_{ki} \} \bigr ] = \eta_{ki}(t) =
\bx_{ki}^\top(t) \bfbeta_k + \bz_{ki}^\top(t) \bb_{ki} \label{Eq:MixedModel}
\end{gather}
\normalsize
where $g_k(\cdot)$ denotes a known one-to-one monotonic link function,  $y_{ki}(t)$ denotes the value of the $k$-th longitudinal outcome for the $i$-th subject at time point $t$, and $\bx_{ki}(t)$ and $\bz_{ki}(t)$ denote the design vectors for the fixed-effects $\bfbeta_k$ and the random effects $\bb_{ki}$, respectively. The dimensionality and composition of these design vectors is allowed to differ between the multiple outcomes, and they may also contain a combination of baseline and time-varying covariates. To account for the association between the multiple longitudinal outcomes we link their corresponding random effects. More specifically, the complete vector of random effects $\bb_i = (\bb_{1i}^\top, \bb_{2i}^\top, \ldots, \bb_{Ki}^\top)^\top$ is assumed to follow a multivariate normal distribution with mean zero and variance-covariance matrix $\bD$.

For the survival process, we assume that the risk for an event depends on a function of the subject-specific linear predictor $\eta_i(t)$ and/or the random effects. More specifically, we have
\small
\begin{gather}
h_i (t \mid \mathcal H_i(t), \bw_i(t)) =\frac{\Lim{\Delta t \rightarrow 0} \Pr \{ t \leq T_i^* < t + \Delta t \mid T_i^* \geq t, \mathcal H_i(t), \bw_i(t) \} }{ \Delta t}, \, t>0 \nonumber \\[2pt]
=  h_0(t)  \exp \biggl  [\bfgamma^\top
\bw_i(t) + \sum \limits_{k = 1}^K \sum \limits_{l = 1}^{L_k} f_{kl} \{\mathcal H_{ki}(t), \bw_i(t), \bb_{ki}, \bfalpha_{kl} \} \biggr], \label{Eq:Surv-RR}
\end{gather}
\normalsize
where $\mathcal H_{ki}(t) = \{ \eta_{ki}(s), 0 \leq s < t \}$ denotes the history of the underlying longitudinal process up to $t$, $h_0(\cdot)$ denotes the baseline hazard function, and $\bw_i(t)$ is a vector of exogenous, possibly time-varying, covariates with corresponding regression coefficients $\bfgamma$. Functions $f_{kl}(\cdot)$, parameterized by vector $\bfalpha_{kl}$, specify which components/features of each longitudinal outcome are included in the linear predictor of the relative risk model \cite{brown:09, rizopoulos.ghosh:11, rizopoulos:12, rizopoulos.et.al:14}. Some examples, motivated by the literature, are (subscripts $kl$ have been dropped in the following expressions but are assumed):
\small
\begin{gather*}
f \{\mathcal H_i(t), \bw_i(t), \bb_i, \bfalpha \}  =  \alpha \eta_i(t),\\
f \{\mathcal H_i(t), \bw_i(t), \bb_i, \bfalpha \}  =  \alpha_1 \eta_i(t) + \alpha_2 \eta_i'(t),
\; \eta_i'(t) = \frac{d\eta_i(t)}{dt},\\
f \{\mathcal H_i(t), \bw_i(t), \bb_i, \bfalpha \}  =  \alpha \int_0^t \eta_i(s) \, ds.
\end{gather*}
\normalsize
These formulations of $f(\cdot)$ postulate that the hazard of an event at time $t$ may be associated with the underlying level of the biomarker at the same time point, the slope of the longitudinal profile at $t$ or the accumulated longitudinal process up to $t$. In addition, the specified terms from the longitudinal outcomes may also interact with some covariates in the $\bw_i(t)$. Furthermore, note, that we allow a combination of $L_k$ functional forms per longitudinal outcome. Finally,  the baseline hazard function $h_0(\cdot)$ is modeled flexibly using a B-splines approach, i.e.,
\small
\begin{gather}
\log h_0(t) = \sum \limits_{q = 1}^Q \gamma_{h_0,q} B_q(t, \bv),
\label{Eq:BaseHaz}
\end{gather}
\normalsize
where $B_q(t, \bv)$ denotes the $q$-th basis function of a B-spline with knots $v_1, \ldots, v_Q$ and $\bfgamma_{h_0}$ the vector of spline coefficients; typically $Q = 15$ or 20.


\section{Likelihood and Priors} 
\label{sec:2}
As explained in Section~\ref{intro}, we use a Bayesian approach for the estimation of the joint model's parameters. The posterior distribution of the model parameters given the observed data is derived under the assumptions that given the random effects, the longitudinal outcomes are independent from the event times, the multiple longitudinal outcomes are independent of each other, and the longitudinal responses of each subject in each outcome are independent. Under these assumptions the posterior distribution is analogous to:
\small
\begin{gather}
p(\bftheta, \bb) \propto \prod_{i = 1}^n \prod_{k = 1}^K \prod_{j = 1}^{n_{ki}} p (y_{kij} \mid \bb_{ki}, \bftheta) \, p(T_i, T_i^U, \delta_i \mid \bb_{ki}, \bftheta) \, p(\bb_{i} \mid \bftheta) \, p(\bftheta), \label{Eq:FullPost}
\end{gather}
\normalsize
where $\bftheta$ denotes the full parameter vector, and
\small
\begin{gather*}
p (y_{kij} \mid \bftheta, \bb_{ki}) = \exp \biggl \{ \frac{\Bigl [y_{kij} \psi_{kij}(\bb_{ki}) - c_k\{\psi_{kij}(\bb_{ki})\} \Bigr ]} {a_k(\varphi) - d_k(y_{kij}, \varphi)} \biggr \},
\end{gather*}
\normalsize
with $\psi_{kij}(\bb_{ki})$ and $\varphi$ denoting the natural and dispersion parameters in the exponential family, respectively, and $c_k(\cdot)$, $a_k(\cdot)$, and $d_k(\cdot)$ are known functions specifying the member of the exponential family. For the survival part accordingly we have
\small
\begin{gather}
p(T_i, T_i^U, \delta_i \mid \bb_i, \bftheta) = \, \Bigr \{h_i (T_i \mid \mathcal H_i(T_i), \bw_i(T_i)) \Bigr \}^{I(\delta_i = 1)} 
\times \exp \biggl \{- \int_0^{T_i} h_i (s \mid \mathcal H_i(s), \bw_i(s)) \; ds \biggr \}^{I(\delta_i \in \{0, 1\})} \nonumber \\[2pt]
\times \Biggl \{1 - \exp \biggl \{- \int_0^{T_i} h_i (s \mid \mathcal H_i(s), \bw_i(s)) \; ds \biggr \}\Biggr\}^{I(\delta_i = 2)} \nonumber \\[2pt]
\times \Biggl \{\exp \biggl \{- \int_0^{T_i} h_i (s \mid \mathcal H_i(s), \bw_i(s)) \; ds \biggr \} 
- \exp \biggl \{- \int_0^{T_i^U} h_i (s \mid \mathcal H_i(s), \bw_i(s)) \; ds \biggr \} \Biggr\}^{I(\delta_i = 3)}, 
\label{Eq:Surv-density}
\hspace{-20pt} 
\raisetag{3\baselineskip}
\end{gather}
\normalsize
where $I(\cdot)$ denotes the indicator function. The integral in the definition of the cumulative hazard function does not have a closed-form solution, and thus a numerical method is employed for its evaluation. Standard options are the Gauss-Kronrod and Gauss-Legendre quadrature rules. 
For the parameters of the longitudinal outcomes we use standard default priors. The covariance matrix of the random effects is parameterized in terms of a correlation matrix $\bfOmega$ and a vector of $\bfsigma_{d}$. For the correlation matrix $\bfOmega$ we use the LKJ-Correlation prior proposed by \cite{lewandowski.et.al:09} with parameter $\zeta = 1.5$. For each element of $\bfsigma_{d}$ we use a half-Student's t prior with 3 degrees of freedom. For the regression coefficients $\bfgamma$ of the relative risk model we assume independent normal priors with zero mean and variance 1000. The same prior is also assumed for the vector of association parameters $\bfalpha$. However, when $\bfalpha$ becomes high dimensional (e.g., when several functional forms are considered per longitudinal outcome), we opt for a global-local ridge-type shrinkage prior. More specifically, for the $s$-th element of $\bfalpha$ we assume
\small
\begin{gather*}
\alpha_s \sim \mathcal N (0, \tau \psi_s), \\[2pt] 
\tau^{-1} \sim \mbox{Gamma}(0.1, 0.1),  \\[2pt] 
\psi_s^{-1} \sim \mbox{Gamma}(1, 0.01) .
\end{gather*}
\normalsize
The global smoothing parameter $\tau$ has sufficient mass near zero to ensure shrinkage, while the local smoothing parameter $\psi_s$ allows individual coefficients to attain large values. The motivation for using this type of prior distribution in this case is that we expect the different terms behind the specification of $f(\cdot)$ to be correlated, and many of the corresponding coefficients to be non-zero. Nonetheless, other options of shrinkage or variable-selection priors could also be used \cite{andrinopoulou.rizopoulos:16}. Finally, the penalized version of the B-spline approximation to the baseline hazard is specified using the following hierarchical prior for $\bfgamma_{h_0}$ \cite{lang.brezger:04}:
\small
\begin{gather*}
p(\bfgamma_{h_0} \mid \tau_h) \propto \tau_h^{\rho(K)/2}\exp \Bigl (-\frac{\tau_{h}}{2}
\bfgamma_{h_0}^\top \bK \bfgamma_{h_0} \Bigr ),
\end{gather*}
\normalsize
where $\tau_h$ is the smoothing parameter that takes a $\mbox{Gamma}(1, \tau_{h\delta} )$ prior distribution, with a hyper-prior $\tau_{h\delta} \sim \mbox{Gamma}(10^{-3}, 10^{-3})$, which ensures a proper posterior distribution for $\bfgamma_{h_0}$ \cite{jullion.lambert:07}, $\bK = \Delta_r^\top \Delta_r + 10^{-6}\bbI$, with $\Delta_r$ denoting the $r$-th difference penalty matrix, and where $\rho(\bK)$ denotes the rank of $\bK$.


\section{Corrected Two-Stage Approach}  \label{sec:3}
\subsection{Importance sampling correction} 
\label{subsec:3.1}
Carrying out a full Bayesian estimation of the multivariate joint model is straightforward, using either Markov chain Monte Carlo (MCMC) or Hamiltonian Monte Carlo (HMC). However, this estimation becomes very challenging from a computational viewpoint, due to the high number of random effects involved, and the requirement for numerical integration in the calculation of the density of the survival outcome (\ref{Eq:Surv-density}). This limitation has hampered the use of multivariate joint models in practice.
The two-stage approach, which entails fitting the longitudinal and survival outcomes separately, is the solution most often used to overcome this computational deadlock. Using this approach, under the Bayesian framework, we would have the following two stages:\\
\begin{enumerate}[leftmargin=0.7cm]
	\item[S-I:] We fit a multivariate mixed model for the longitudinal outcomes using either MCMC or HMC, and we obtain a sample $\{\bftheta_y^{(m)}, \bb^{(m)}; m = 1, \ldots, M\}$ of size $M$ from the posterior,
\end{enumerate}
\small
\begin{gather*}
p(\bftheta_y, \bb \mid \by) \propto \prod \limits_{i = 1}^n \prod \limits_{k = 1}^K \prod \limits_{j = 1}^{n_{ki}} p (y_{kij} \mid \bb_{ki}, \bftheta) \; p(\bb_{i} \mid \bftheta) \; p(\bftheta_y),
\end{gather*}
\normalsize
where $\bftheta_y$ denotes the subset of the parameters that are included in the definition of the longitudinal submodels (including the parameters in the random-effects distribution).
\begin{enumerate}[leftmargin=0.7cm]
	\item[S-II:] Utilizing the sample from Stage I, we obtain a sample for the parameters of the survival submodel\\ $\{\bftheta_t^{(m)}; m = 1, \ldots, M\}$ from the corresponding posterior distribution,
\end{enumerate}
\small
\begin{gather*}
p(\bftheta_t \mid \tilde T, \delta, \bftheta_y^{(m)}, \bb^{(m)}) \propto  \prod \limits_{i = 1}^n p(\tilde T_i, \delta_i \mid \bftheta_t, \bb_i^{(m)}, \bftheta_y^{(m)}) \; p(\bftheta_t),
\end{gather*}
\normalsize
where $\bftheta_t$ denotes the subset of the parameters that are included in the definition of the survival submodel, and $\tilde T = (T, T^U)$.
This two-stage procedure essentially entails the same number of iterations as the full Bayesian estimation of the multivariate joint model. The computational benefits stem from the fact that we do not need to numerically integrate the survival submodel density function in Stage I. 
Even though this approach greatly reduces the computational burden, there exists a substantial body of work demonstrating that it results in biased estimates, even in the simpler case of univariate joint models \cite[see][and references therein]{tsiatis.davidian:04, rizopoulos:12}. This bias is a result of not working with the full joint distribution, which would produce estimates of $\bftheta_y$ and $\bb$ that are appropriately corrected for informative dropout relating to the occurrence of an event.

To overcome this issue, we propose the correction of the estimates we obtain from the two-stage approach using importance sampling weights \cite[Section~7.9]{press.et.al:07}. In particular, we consider that the realizations $\{\bftheta_t^{(m)}, \bftheta_y^{(m)}, \bb^{(m)}; m = 1, \ldots, M\}$ that we have obtain using the two-stage approach can be considered a weighted sample from the full posterior of the multivariate joint model with weights given by:
\small
\begin{gather}
w^{(m)} = \frac{p(\bftheta_t^{(m)} \mid \tilde T, \delta, \bftheta_y^{(m)}, \bb^{(m)}) \; p(\bftheta_y^{(m)}, \bb^{(m)} \mid \by, \tilde T, \delta)}{p(\bftheta_t^{(m)} \mid \tilde T, \delta, \bftheta_y^{(m)}, \bb^{(m)}) \; p(\bftheta_y^{(m)}, \bb^{(m)} \mid \by)}. \label{Eq:ISweights}
\end{gather}
\normalsize
The numerator in this expression is the posterior distribution of the multivariate joint model, and the denominator, the corresponding posterior distributions from each of the two stages. As previously stated, from (\ref{Eq:ISweights}) we observe that the difference between fitting the full joint model versus the two-stage approach comes from the second term in the numerator and denominator. By expanding these two terms we obtain
\small
\begin{gather}
\frac{p(\bftheta_y^{(m)}, \bb \mid \by, \tilde T, \delta)}{p(\bftheta_y^{(m)}, \bb^{(m)} \mid \by)}  \propto 
\frac{\prod_i p(\by_i \mid \bb_i^{(m)}, \bftheta_y^{(m)}) \, p(\tilde T_i, \delta_i \mid \bb_i^{(m)}, \bftheta_y^{(m)}) \, p(\bb_i^{(m)} \mid \bftheta_y^{(m)}) \, p(\bftheta_y^{(m)})}{\prod_i p(\by_i \mid \bb_i^{(m)}, \bftheta_y^{(m)}) \, p(\bb_i^{(m)} \mid \bftheta_y^{(m)})  
	\, p(\bftheta_y^{(m)})} \nonumber \\[2pt]
= \prod_i p(\tilde T_i, \delta_i \mid \bb_i^{(m)}, \bftheta_y^{(m)}) \nonumber \\[2pt] 
= \prod_i \int p(\tilde T_i, \delta_i \mid \bb_i^{(m)}, \bftheta_y^{(m)}, \bftheta_t) \, d\bftheta_t. \label{Eq:ISweights-worked}
\end{gather}
\normalsize
The resulting weights involve a marginal likelihood calculation, which we perform using a Laplace approximation, namely
\small
\begin{gather*}
\varpi^{(m)}  =  \exp \Bigl [ \frac{q \log(2 \pi) - \log\{\det(\widehat{\Sigma}^{(m)})\}}{2}  
+ \log \{p(\tilde T_i, \delta_i \mid \bb_i^{(m)}, \bftheta_y^{(m)}, \widehat{\bftheta}_t^{(m)})\} \Bigr],
\\[2pt]
\tilde w^{(m)}  =  \varpi^{(m)} \Big / \sum_{m = 1}^M \varpi^{(m)},
\end{gather*}
\normalsize
where 
\small 
\begin{gather*}
\widehat{\bftheta}_t^{(m)} = \argmax_{\bftheta_t} \bigl [ \log \{p(\tilde T_i, \delta_i \mid \bb_i^{(m)}, \bftheta_y^{(m)}, \widehat{\bftheta}_t)\} \bigr],
\end{gather*}
\normalsize 
$\det(A)$ denotes the determinant of matrix $A$, 
\small 
\begin{gather*}
\widehat{\Sigma}^{(m)} = - \partial^2 \log \{p(\tilde T_i, \delta_i \mid \bb_i^{(m)}, \bftheta_y^{(m)}, \widehat{\bftheta}_t)\} \big / \partial \bftheta_t^\top \partial \bftheta_t \big |_{\theta_t = \hat{\theta}_t^{(m)}},
\end{gather*}
\normalsize 
and $q$ denotes the dimensionality of the $\bftheta_t$ vector. The extra computational burden of performing this Laplace approximation is minimal in practice, since good initial values can be provided from one iteration $m$ to the next $m + 1$, which substantially reduces the number of required optimization iterations for finding $\widehat{\bftheta}_t^{(m)}$ (i.e., $\widehat{\bftheta}_t^{(m)}$ is provided as an initial value to find $\widehat{\bftheta}_t^{(m + 1)}$).

\subsection{Performance} 
\label{subsec:3.2}
To evaluate whether the introduction of the importance sampling weights alleviates the bias observed with the simple two-stage approach (i.e., without the weights), we perform a `proof-of-concept' simulation study. In particular, we compare the proposed corrected two stage approach with the simple two-stage approach, as well as the full multivariate joint model in the case of two continuous longitudinal outcomes. The specific details of this simulation setting are given in Appendix~\ref{appendix:simI}. The results from 500 simulated datasets are presented in Figure~\ref{fig:1} and in the appendix, in Figures~\ref{fig:4} and~\ref{fig:5}.
\begin{figure}
	\begin{center}
		\includegraphics[width = \columnwidth]{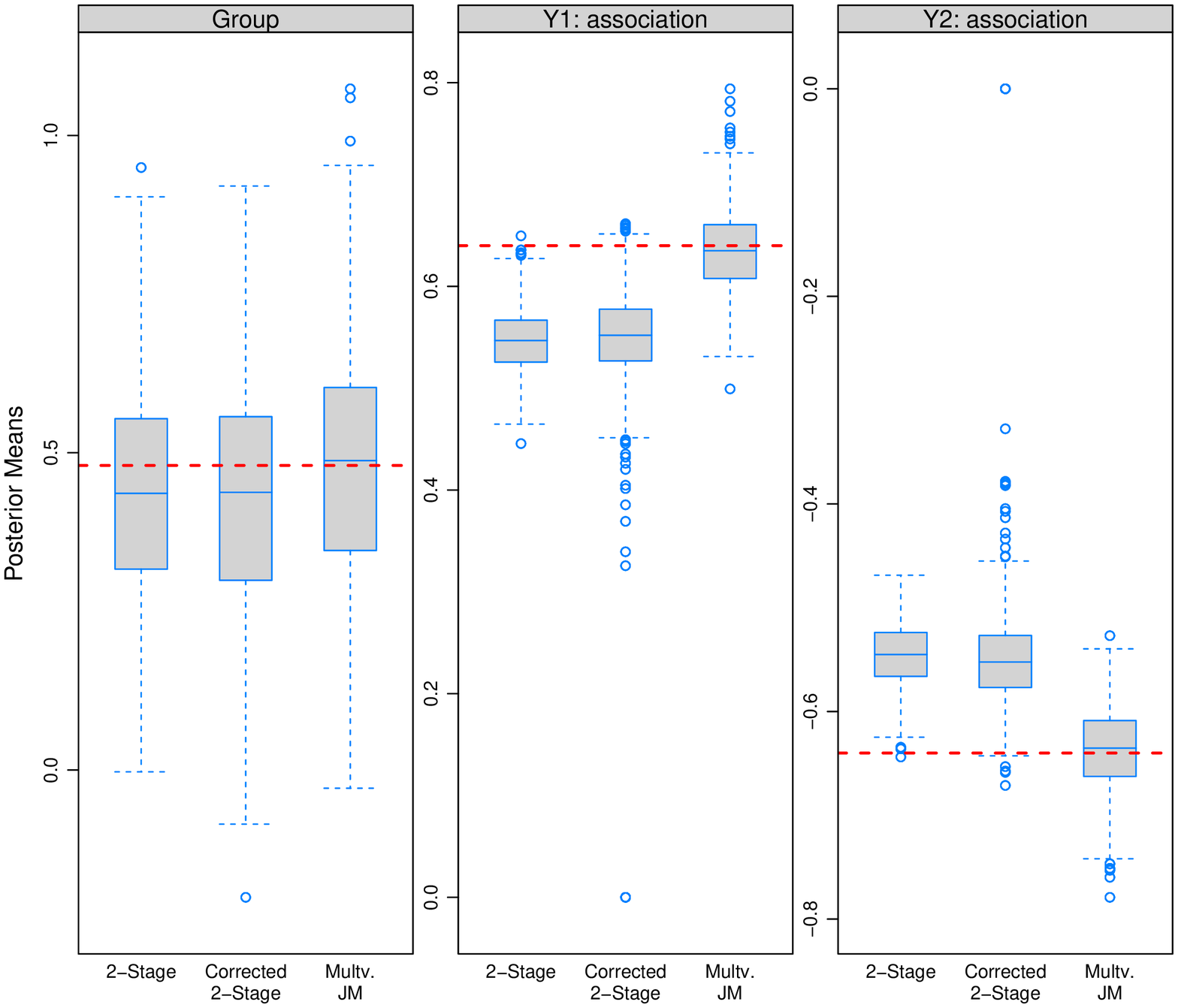}
	\end{center}
	\caption{Simulation results from 500 datasets comparing the two-stage approach and the importance-sampling-corrected two-stage approach with the full joint model for continuous longitudinal outcomes. The three panels show posterior means from the 500 datasets for the three coefficients in the survival submodel, namely the coefficient for the baseline group variable and the association parameters for the two longitudinal outcomes. The dashed horizontal line indicates the true value of the coefficients.} 
	\label{fig:1}
\end{figure}
Figure~\ref{fig:4} shows boxplots with the computing times required to fit the joint model under three approaches. Comparing the first two of these approaches, we see that the calculation of the importance-sampling weights in the corrected two-stage approach had minimal computational cost, with the full multivariate joint model taking substantially more time to fit. Figure~\ref{fig:5} shows boxplots of posterior means from the 500 datasets for the parameters of the two longitudinal submodels. We observe that all three approaches provide very similar results with minimal bias. Figure~\ref{fig:1} shows the corresponding boxplots of posterior means for the parameters of the survival submodel. As expected, the full multivariate joint model returns unbiased results. Similarly, as has previously been reported in the literature, the simple two-stage approach exhibits considerable bias. We see that this bias persists for the corrected two-stage approach, although theoretically the use of the importance sampling weights should alleviate it (by adjusting the posterior means obtained via the simple two-stage approach such that they more closely resemble those from the full multivariate model).

\section{Corrected Two-Stage Approach with Random Effects}  \label{sec:4}
\subsection{Importance sampling correction with random effects}
\label{subsec:4.1}
\begin{figure}
	\begin{center}
		\includegraphics[angle = -90, width = \columnwidth]{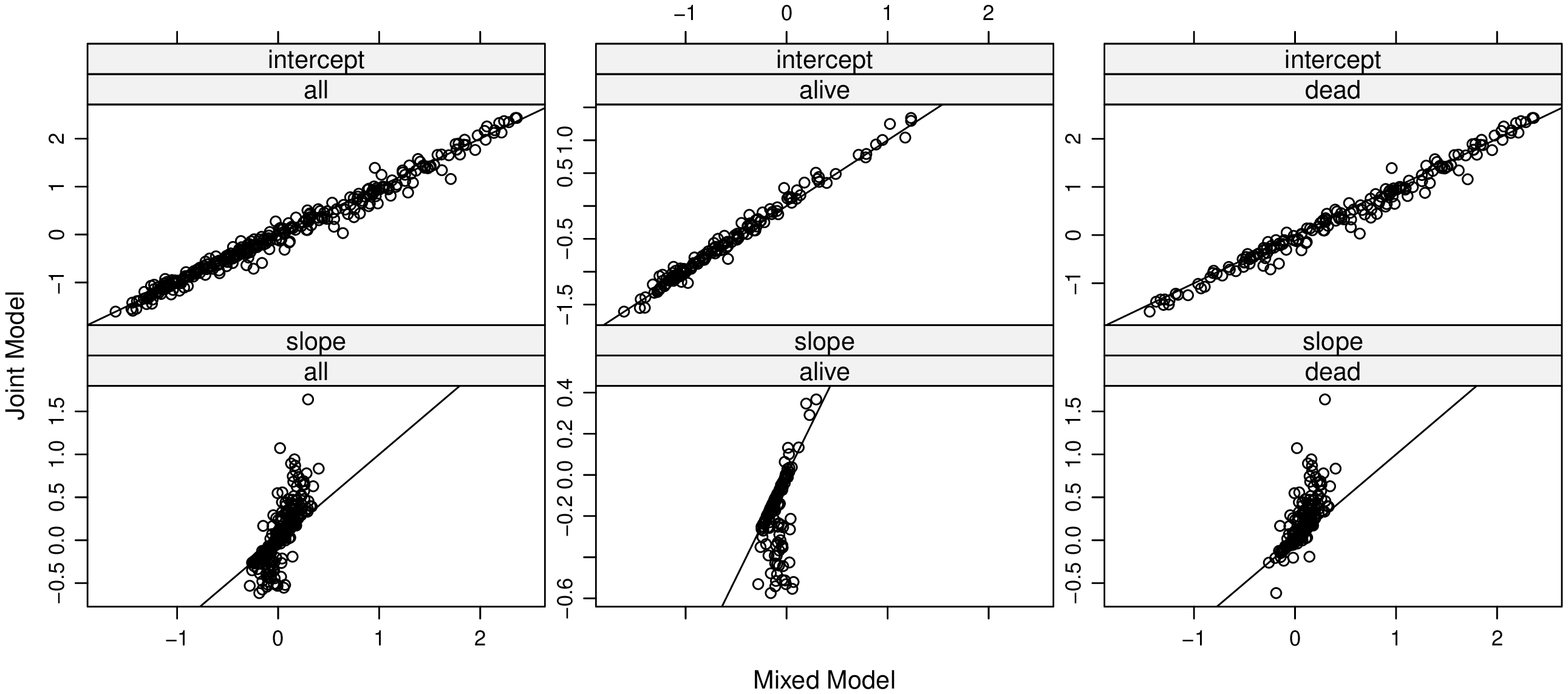}
	\end{center}
	\caption{Comparison of posterior mean estimates for the random intercepts and random slopes from one simulated dataset for the first longitudinal outcome between a linear mixed model and a joint model. The left column panels correspond to all subjects, the middle column to subjects without an event, and the right column panel to subjects with an event.} \label{fig:2}
\end{figure}

The above result is unexpected, since (as per Figure~\ref{fig:5}), the corrected two-stage (and indeed the simple two-stage) approach unbiasedly estimates both the fixed effects and the variance components of the longitudinal submodels. However, further investigation shows that there is a considerable difference between the corrected two-stage approach and the multivariate joint model with regards to the posterior of the random effects. This is depicted in Figure~\ref{fig:2} for one of the longitudinal outcomes we have simulated.
The data have been simulated such that higher values for longitudinal outcome $y_1$ are associated with a higher hazard of the event. From Figure~\ref{fig:2} we observe that the random effect estimates for the multivariate mixed model, and especially the random slope estimates for subjects with and without an event differ from those for the multivariate joint model. In particular, we observe that the random slope estimates from the joint model are larger for subjects with an event compared to the linear mixed model, and vice versa for subjects without an event. This observation suggests that we could improve the weights given in (\ref{Eq:ISweights}) by updating (in the second stage) not only the parameters of the survival submodel $\bftheta_t$ but also the random effects $\bb$. That is, we obtain a sample for the parameters of the survival submodel $\{\bftheta_t^{(m)}, \bb^{(m)}; m = 1, \ldots, M\}$ from the corresponding joint posterior distribution,
\small
\begin{gather}
p(\bftheta_t, \bb \mid \tilde T, \delta, \by, \bftheta_y^{(m)}) \propto
\prod \limits_{i = 1}^n \prod \limits_{k = 1}^K \prod \limits_{j = 1}^{n_{ki}} p (y_{kij} \mid \bb_{ki}, \bftheta_y^{(m)}) \; p(\bb_{i} \mid \bftheta_y^{(m)})  \, p(\tilde T_i, \delta_i \mid \bftheta_t, \bb_i, \bftheta_y^{(m)}) \; p(\bftheta_t). \label{Eq:Second_Stage_RE}
\end{gather}
\normalsize
Admittedly, simulating from $[\bftheta_t, \bb \mid \tilde T, \delta, \by, \bftheta_y^{(m)}]$ is more computationally intensive than simulating from $[\bftheta_t \mid \tilde T, \delta, \bftheta_y^{(m)}, \bb^{(m)}]$, the corresponding second stage presented in Section~\ref{sec:3}, since we now also need to calculate the densities of the mixed-effect models for the $K$ longitudinal outcomes. Nonetheless, the computational gains compared to fitting the full joint model remain significant.
Under this second stage (\ref{Eq:Second_Stage_RE}) the importance sampling weights now take the form:
\small
\begin{gather}
w^{(m)} = \frac{p(\bftheta_t^{(m)}, \bb^{(m)} \mid \tilde T, \delta, \bftheta_y^{(m)}) \; p(\bftheta_y^{(m)} \mid \by, \tilde T, \delta)}{p(\bftheta_t^{(m)}, \bb^{(m)} \mid \tilde T, \delta, \by, \bftheta_y^{(m)}) \; p(\bftheta_y^{(m)}, \bb^{(m)} \mid \by)}. \label{Eq:ISweights_RE}
\end{gather}
\normalsize
Similarly to (\ref{Eq:ISweights}), the new weights have been formulated such that the difference lies in the second term in both the numerator and denominator. By doing an expansion of these two terms similar to that used in the previous section, we obtain:
\small
\begin{gather}
\nonumber w^{(m)} = \frac{p(\bftheta_y^{(m)} \mid \by, \tilde T, \delta)}{p(\bftheta_y^{(m)}, \bb^{(m)} \mid \by)} \nonumber \\ \propto \varpi^{(m)}  
= \frac{\prod_i p(\by_i, \tilde T_i, \delta_i \mid \bftheta_y^{(m)}) \; p(\bftheta_y^{(m)})}{\prod_i p(\by_i \mid \bb_i^{(m)}, \bftheta_y^{(m)}) \; p(\bb_i^{(m)} \mid \bftheta_y^{(m)}) \; p(\bftheta_y^{(m)})}\nonumber \\
=  \frac{\prod_i \displaystyle \int \int p(\by_i \mid \bb_i, \bftheta_y^{(m)}) p(\tilde T_i, \delta_i \mid \bb_i, \bftheta_y^{(m)}, \bftheta_t) \, p(\bb_i \mid \bftheta_y^{(m)}) \, p(\bftheta_t) \, d\bb_i d\bftheta_t}{\prod_i p(\by_i \mid \bb_i^{(m)}, \bftheta_y^{(m)}) \; p(\bb_i^{(m)} \mid \bftheta_y^{(m)})}, \label{Eq:ISweights_RE-worked}
\end{gather}
\normalsize
and the self-normalized weights are 
\small
\begin{gather*}
\tilde w^{(m)} = \varpi^{(m)} / \sum_m \varpi^{(m)}.
\end{gather*} 
\normalsize
The integrals in the numerator are once again approximated using the Laplace method, namely, we let
\small
\begin{gather*}
\bigl \{ \widehat{\bftheta}_t^\top, \widehat{\bb}_i^\top \bigr \}^\top 
=  \argmax_{\theta_t, b_i} \Bigl \{ \sum_j \log p(y_{ij} \mid \bb_i, \bftheta_y^{(m)})  +  \log p(\tilde T_i, \delta_i \mid \bb_i, \bftheta_y^{(m)}, \bftheta_t) + \log p(\bb_i \mid \bftheta_y^{(m)}) + \log p(\bftheta_t) \Bigr \},
\end{gather*}
\normalsize
and
\small
\begin{gather*}
\Sigma_{b_i} = - \frac{\partial^2 \bigl \{ \log p(\by_i \mid \bb_i, \bftheta_y^{(m)}) + \log p(\tilde T_i, \delta_i \mid \bb_i, \bftheta_y^{(m)}, \widehat{\bftheta}_t) + \log p(\bb_i \mid \bftheta_y^{(m)}) \bigr \} }{\partial \bb^\top \partial \bb} \Big |_{b = \hat b_i},
\end{gather*}
\normalsize
denote the Hessian matrix for the random effects, and analogously,
\small
\begin{gather*}
\Sigma_{\theta_t} = - \frac{\partial^2 \sum_i \bigl \{ \log p(\tilde T_i, \delta_i \mid \widehat{\bb}_i, \bftheta_y^{(m)}, \bftheta_t) + \log p(\bftheta_t) \bigr \} }{\partial \bftheta_t^\top \partial \bftheta_t} \Big |_{\theta_t = \hat \theta_t},
\end{gather*}
\normalsize
denote the Hessian matrix for the $\bftheta_t$ parameters. Then, we approximate the inner integral by
\small
\begin{gather*}
p(\by_i, \tilde T_i, \delta_i \mid \bftheta_y^{(m)}, \widehat{\bftheta}_t)
\approx  \exp \biggl [\frac{\kappa \log(2\pi) - \log \bigl \{ \mbox{det}(\Sigma_{b_i}) \bigr \} }{2} + \log p(\by_i \mid \widehat{\bb}_i, \bftheta_y^{(m)}) + \log p(\tilde T_i, \delta_i \mid \widehat{\bb}_i, \bftheta_y^{(m)}, \widehat{\bftheta}_t) + \\
\hspace*{2cm} \log p(\widehat{\bb}_i \mid \bftheta_y^{(m)}) \biggr ],
\end{gather*}
\normalsize
where $\kappa$ denotes the number of random effects for each subject $i$. Similarly, the outer integral is approximated as
\small
\begin{gather*}
p(\by_i, \tilde T_i, \delta_i \mid \bftheta_y^{(m)}) \approx \exp \biggl [\frac{q \log(2\pi) - \log \bigl \{ \mbox{det}(\Sigma_\theta) \bigr \} }{2} +  \sum_i \log p(\by_i, \tilde T_i, \delta_i \mid \bftheta_y^{(m)}, \widehat{\bftheta}_t) \biggr ].
\end{gather*}
\normalsize
Given the requirement for a double Laplace approximation, and the fact that the denominator does not simplify, the calculation of the $\varpi^{(m)}$ weights given by (\ref{Eq:ISweights_RE-worked}) is more computationally intensive than the ones presented in Section~\ref{sec:3}. Nevertheless, these required computations still remain many orders of magnitude faster than fitting the full joint model.

\subsection{Performance}
\label{subsec:4.2}
\begin{figure}
	\begin{center}
		\includegraphics[width = \columnwidth]{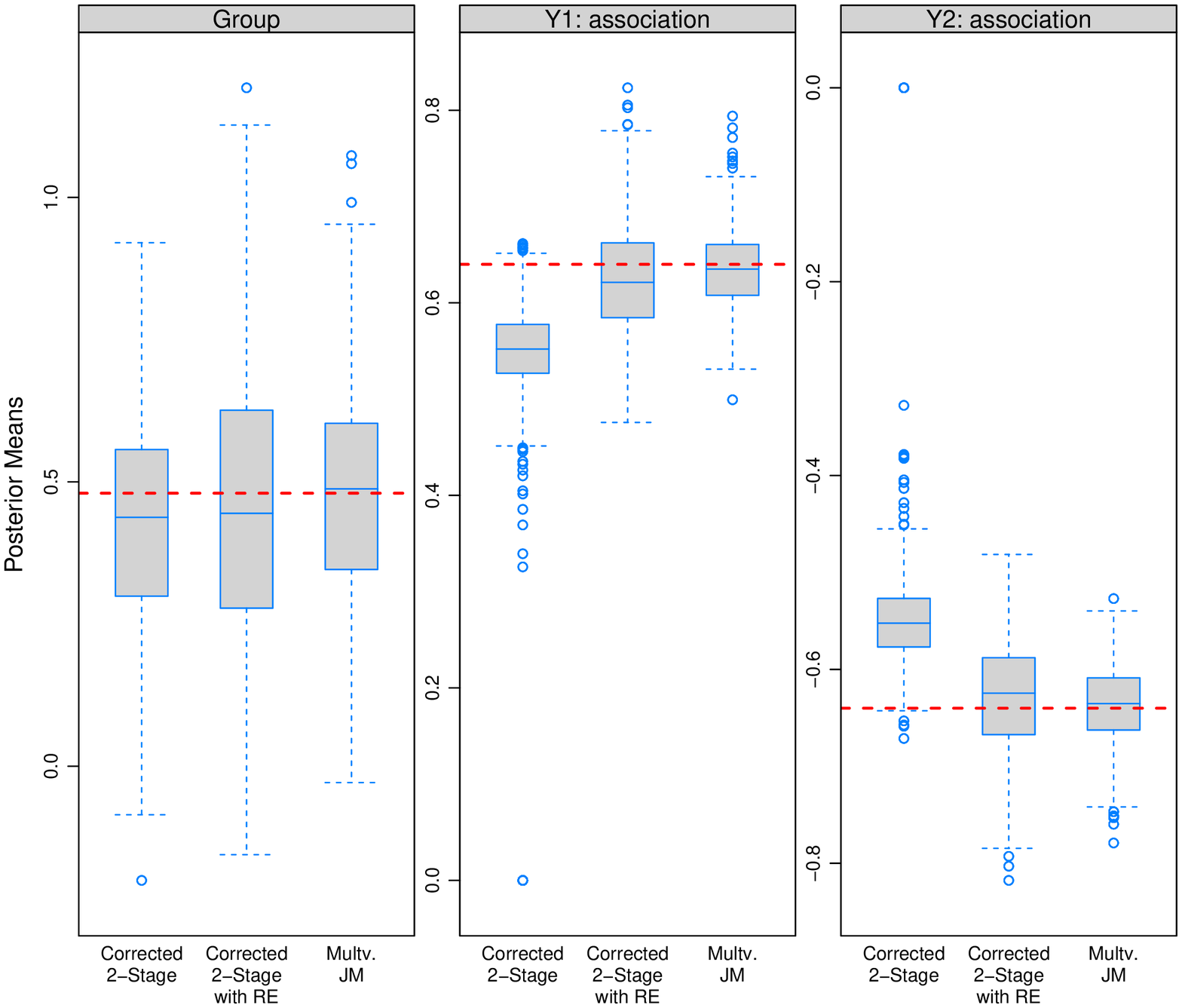}
	\end{center}
	\caption{Simulation results from 500 datasets comparing the importance-sampling-corrected two-stage approach and the corrected two-stage approach with random effects with the full joint model for continuous longitudinal outcomes. The three panels show posterior means from the 500 datasets for the three coefficients in the survival submodel, namely the coefficient for the baseline group variable and the association parameters for the two longitudinal outcomes. The dashed horizontal line indicates the true value of the coefficients.} \label{fig:3}
\end{figure}

To assess whether updating the random effects in the importance sampling weights alleviates the bias we observed in Section~\ref{subsec:3.2}, we have re-analyzed the same simulated datasets. The details are again given in Appendix~\ref{appendix:simI}. The results from 500 simulated datasets are presented in Figures~\ref{fig:3},~\ref{fig:4}, and~\ref{fig:6}.
As anticipated, the corrected two-stage approach with updated random effects added only a small computational cost, with the full multivariate joint model still taking considerably more time to fit than either of the corrected two-stage approaches (Figure~\ref{fig:4}). The boxplots depicting the posterior means from the 500 datasets for the parameters of the longitudinal submodels once again demonstrate similar results for all three approaches (Figure~\ref{fig:6}). Figure~\ref{fig:3} shows the posterior means for the parameters of the survival submodel. We observe that the bias seen for the corrected two-stage approach is now eliminated, with the posterior means from the approach with updated random effects closely approximating those from the full multivariate joint model. 
 

\section{Extra Simulations} 
\label{sec:5}
Further simulations were performed in order to assess the performance of the importance-sampling-corrected two-stage approach with the updated random effects, in different scenarios. Details of these simulations are given in Appendices~\ref{appendix:simII} and~\ref{appendix:simIII}.

\subsection{Scenario II}\label{subsec:5.1}
Scenario II included 6 continuous longitudinal outcomes. Owing to the increased number of outcomes, the full multivariate joint model was not run. Table~\ref{tab:1} shows the bias for the parameters of the survival submodel, together with the RMSE and coverage (based on the 2.5\% and 97.5\% credibility intervals for each parameter). Table~\ref{tab:4} in the appendix shows the same information for the parameters of the 6 longitudinal outcomes. 


\begin{table}
	\begin{center}
	\begin{tabular}{lrrrr}
		\hline\noalign{\smallskip}
		Parameter	& True Value &	Bias	&	RMSE	&	Coverage	\\
		
		\noalign{\smallskip}\hline\noalign{\smallskip}
Group	&	-0.20 &	-0.02	&	0.19	&	0.96	\\
$\alpha_1$	& 1.09	&	-0.01	&	0.04	&	0.93	\\
$\alpha_2$	&	-1.09	&	0.01	&	0.04	&	0.93	\\
$\alpha_3$	&	-0.92	&	-0.01	&	0.16	&	0.95	\\
$\alpha_4$	&	0.92	&	0.03	&	0.17	&	0.95	\\
$\alpha_5$	&	0.43	&	0.03	&	0.05	&	0.83	\\
$\alpha_6$	&	-0.43	&	-0.02	&	0.04	&	0.88	\\
		
		\noalign{\smallskip}\hline
	\end{tabular}
\caption{Simulation results for parameters of the survival submodel (Scenario II)}
\label{tab:1}       
\end{center}
\end{table}

\subsection{Scenario III}\label{subsec:5.2}
Scenario III again included 6 longitudinal outcomes, now of varying types: 3 continuous and 3 binary. Table~\ref{tab:2} demonstrates yet again the alleviation of the bias achieved by updating the random effects.


\begin{table}
	\begin{center}
	\begin{tabular}{lrrrr}
		\hline\noalign{\smallskip}
			Parameter	& True Value &	Bias	&	RMSE	&	Coverage	\\
		\noalign{\smallskip}\hline\noalign{\smallskip}
		Group	& -0.47& 	-0.01	&	0.91	&	0.97	\\
		$\alpha_1$	& 0.17 & 	-0.01	&	0.04	&	1.00	\\
		$\alpha_2$	& -0.17& 	0.01	&	0.04	&	1.00	\\
		$\alpha_3$	& 0.17&	0.00	&	0.43	&	0.90	\\
		$\alpha_4$	& -0.17& 	0.02	&	0.88	&	0.96	\\
		$\alpha_5$	& 0.47& 	0.02	&	0.04	&	1.00	\\
		$\alpha_6$	& -0.47& 	-0.01	&	0.05	&	0.99	\\
		
		\noalign{\smallskip}\hline
	\end{tabular}
\caption{Simulation results for parameters of the survival submodel (Scenario III)}
\label{tab:2}       
\end{center}
\end{table}


\section{Analysis of the Bio-SHiFT Dataset} 
\label{sec:6}
In this section, we present the analysis of data from the Bio-SHiFT cohort study. During a median follow-up period of 2.4 years (IQR: 2.32  - 2.45), estimated using the reverse Kaplan-Meier methodology \cite{rev_km}, 66/254 (26\%) patients experienced the primary event of interest (a composite event, consisting of hospitalisation for heart failure, cardiac death, LVAD placement and heart transplantation). Biomarkers were measured at inclusion and subsequently every 3 months until the end of follow-up. We focus on 6 biomarkers: the glomerular marker Cystatin C (CysC), two tubular markers; urinary N-acetyl-beta-D-glucosaminidase (NAG) and kidney-injury-molecule (KIM)-1, and the markers N-terminal propBNP (NT-proBNP), cardiac troponin T (HsTNT), and C-reactive protein (CRP). The latter three markers are known to be related to poor outcomes in CHF patients, and measure various aspects of heart failure pathophysiology (wall stress, myocyte damage and inflammation respectively). All biomarkers were logarithmically transformed for further analysis (log base 2) due to skewness. 
For each of NT-probnp, HsTNT and CRP, we included natural cubic splines in both the fixed and random effects parts of their longitudinal models, with differing numbers of knots per outcome (Figure~\ref{fig:7}). Simple linear models with random intercept and slope were used for CysC, NAG and KIM-1. 
Thus, for each of CysC, NAG and KIM-1 ($k = 1, 2 , 3$), we fit:
\small
\begin{gather*}
E\{y_{ki}(t) \mid \mathbf{b}_{ki}\} = \beta_{k0} + b_{ki0}+(\beta_{k1} + b_{ki1})\times \textrm{time}.
\end{gather*}
\normalsize
For the remaining outcomes ($k = 4, 5, 6$), we have:
\small
\begin{gather*}
E\{y_{ki}(t) \mid \mathbf{b}_{ki}\}  =  (\beta_{k0} + b_{ki0}) + \sum_p (\beta_{kp} + b_{kip}) B_{kn}(t,\lambda_p),
\end{gather*}
\normalsize
where $B_{kn}(t; \lambda_p)$ denotes the B-spline basis matrix for a natural cubic spline of time with two internal knots placed at the 25th and 75th percentiles of the follow up times for NT-probnp ($p = 1, 2, 3$), and one internal knot placed at the 50th percentile of the follow up times for each of HsTNT and CRP ($p = 1, 2$). Boundary knots were set at the 5th and 95th percentiles. We assume a multivariate normal distribution for the random effects, $\textbf{b}_i = (\textbf{b}_{1i}^T, \textbf{b}_{2i}^T, \ldots,  \textbf{b}_{6i}^T)^T \sim MVN(\mathbf{0},\textbf{\emph{D}})$, where $\textbf{\emph{D}}$ is a $16 \times 16$ unstructured variance covariance matrix. 
For the survival process, we included the baseline variables: (standardized) age, sex, NYHA class (class III / IV vs. class I / II), use of diuretics, presence or absence of ischemic heart disease (IHD), diabetes mellitus, (standardized) BMI, and the estimated glomerular filtration rate (eGFR) value.

We fit 3 joint models, using the global-local ridge-type shrinkage prior previously described in each case. Model 1 included only the current underlying value of the longitudinal marker for each of the 6 markers. Model 2 included the current value and slope for each marker, and model 3 included the integrated longitudinal profile for each marker (AUC). We thus have:
\small
\begin{gather*}
\textrm{Model 1:} \quad \nonumber h_i (t) = 
h_0(t) \exp \biggl  [\bfgamma^\top
\bw_i(t) + \sum \limits_{k = 1}^K \alpha_k \eta_{ki}(t)\biggr], \\
\textrm{Model 2:} \quad \nonumber h_i (t)  = 
h_0(t) \exp \biggl  [\bfgamma^\top
\bw_i(t) +  \sum \limits_{k = 1}^K \alpha_{1k} \eta_{ki}(t) 
+ \sum \limits_{k = 1}^K \alpha_{2k} \eta_{ki}'(t)\biggr],\\
\textrm{Model 3:} \quad \nonumber h_i (t)  = 
h_0(t) \exp \biggl  [\bfgamma^\top
\bw_i(t) + \sum \limits_{k = 1}^K \alpha_k \int_0^t \eta_{ki}(s) \, ds \biggr].
\label{Eq:Surv-cardio}
\end{gather*}
\normalsize
The parameter estimates and 95\% credibility intervals for the event process are presented in Table~\ref{tab:3}. Hazard ratios are presented per doubling of level, slope or AUC at any point in time. 
\begin{table}

	\begin{center}
		\scalebox{0.8}{\begin{tabular}{lrrr}
				\hline\noalign{\smallskip}
			\multicolumn{4}{c}{Event Process} \\ 
			\noalign{\smallskip}\hline\noalign{\smallskip}	
			&	\multicolumn{3}{c}{Model 1: Current value}		\\	 
			\noalign{\smallskip}\hline\noalign{\smallskip}
			&	HR (95\% CI) & HR*	&	p-value	 	\\ 
			\noalign{\smallskip}\hline\noalign{\smallskip}
			Age	&	1.01 (0.99 to 1.03)	&	1.00	&	0.242	 \\	
			Sex (Male vs. Female)	&	1.00 (0.77 to 1.31)	&	1.06	&	0.970\\	
			NYHA Class (III / IV vs. I / II)&	1.47 (0.97 to 2.64)	&	2.52& 0.134\\
			Diuretics (Yes vs. No)	&	1.12 (0.79 to 2.74)	&	0.96	&	0.770	\\	
			IHD (Yes vs. No)	&	1.06 (0.88 to 1.55)	&	0.95	&	0.696\\
			eGFR	&	1.01 (1.00 to 1.02)	&	1.01	&	0.048	\\		
			BMI 	&	1.04 (0.99 to 1.10)	&	1.03	&	0.140	\\	
			Diabetes Mellitus	&	1.08 (0.86 to 1.79)	&	0.93	&	0.638	\\
			$\alpha_{CRP}$	&	1.44 (1.10 to 1.89)	&	1.25	&	0.008	\\
			$\alpha_{HsTNT }$	&	1.32 (0.99 to 1.84)	&	1.35	&	0.078\\
			$\alpha_{NT-proBNP }$ 	&	1.86 (1.45 to 2.37)	&	2.20	&	$<$0.0001\\
			$\alpha_{CysC }$	&	1.04 (0.55 to 2.48)	&	0.90	&	0.966	\\	
			$\alpha_{NAG }$	&	0.89 (0.51 to 1.38)	&	0.55	&	0.660	\\
			$\alpha_{KIM-1 }$	&	0.96 (0.74 to 1.22)	&	1.21	&	0.768	 \\ 
			\noalign{\smallskip}\hline\noalign{\smallskip}	
			
			&	\multicolumn{3}{c}{Model 2: Current value and slope} 	\\
			\noalign{\smallskip}\hline\noalign{\smallskip}
			& 	HR (95\% CI)  & HR* &		p-value		  \\ 	\noalign{\smallskip}\hline\noalign{\smallskip}
			
			Age	&	1.01 (0.99 to 1.03)	&	1.00	&	0.162	 \\
			Sex (Male vs. Female)	&	0.99 (0.66 to 1.34)	&	0.91	&	0.994		\\
			NYHA Class (III / IV vs. I / II)	&	1.61 (0.97 to 2.97)	&	1.53	&	0.104	\\
			Diuretics (Yes vs. No)	&	1.20 (0.79 to 4.64)	&	0.95	&	0.736	\\
			IHD (Yes vs. No)	&	1.06 (0.84 to 1.56)	&	1.36	&	0.750		\\
			eGFR	&	1.01 (1.00 to 1.02)	&	1.01	&	0.090		\\
			BMI 	&	1.04 (0.98 to 1.10)	&	1.08	&	0.236		\\
			Diabetes Mellitus	&	1.08 (0.84 to 1.73)	&	1.03	&	0.676		\\
			$\alpha_{CRP \, value}$	&	1.52 (1.19 to 1.99)	&	1.38	&	0.002		\\
			$\alpha_{HsTNT \, value}$	&	1.37 (1.00 to 1.89)	&	1.08	&	0.048		\\
			$\alpha_{NT-proBNP \, value}$ 	&	1.90 (1.48 to 2.44)	&	1.61	&	$<$0.0001 	\\
			$\alpha_{CysC \, value}$	&	1.05 (0.47 to 3.20)	&	0.92	&	0.996		\\
			$\alpha_{NAG \, value}$	&	0.74 (0.36 to 1.22)	&	1.04	&	0.292		\\
			$\alpha_{KIM-1 \, value}$	&	0.92 (0.69 to 1.18)	&	1.02	&	0.534		\\
			$\alpha_{CRP \, slope}$	&	0.95 (0.55 to 1.50)	&	0.91	&	0.878		\\
			$\alpha_{HsTNT \, slope}$	&	0.84 (0.24 to 1.91)	&	1.05	&	0.734		\\
			$\alpha_{NT-proBNP \, slope}$ 	&	1.02 (0.70 to 1.48)	&	0.87	&	0.926		\\
			$\alpha_{CysC \, slope}$	&	1.75 (0.19 to 356.56)	&	1.35	&	0.826		\\
			$\alpha_{NAG \, slope}$	&	1.03 (0.08 to 6.63)	&	0.84	&	0.946		\\
			$\alpha_{KIM-1 \, slope}$	&	2.37 (0.61 to 63.48)	&	0.93	&	0.460		 \\ 
			\noalign{\smallskip}\hline\noalign{\smallskip}
			
			& \multicolumn{3}{c}{Model 3: AUC} 		\\
			\noalign{\smallskip}\hline\noalign{\smallskip}
			& 	HR (95\% CI)  & HR* &		p-value		  \\ 	\noalign{\smallskip}\hline\noalign{\smallskip}	
			Age &	1.00 (0.98 to 1.03)	&	1.01	&	0.812  	\\
			Sex &	1.02 (0.83 to 1.38)	&	1.00	&	0.944	\\	
			NYHA Class (III / IV vs. I / II)&	1.80 (0.99 to 3.25)	&	1.83	&	0.080	\\	
			Diuretics (Yes vs. No)&	1.18 (0.83 to 3.72)	&	1.22	&	0.680	\\	
			IHD (Yes vs. No)&	1.04 (0.85 to 1.40)	&	1.21	&	0.742	\\
			eGFR&	1.01 (1.00 to 1.02)	&	1.00	&	0.166	\\
			BMI&	1.03 (0.98 to 1.08)	&	1.02	&	0.286	\\
			Diabetes Mellitus&	1.11 (0.89 to 1.87)	&	0.99	&	0.564	\\
			$\alpha_{CRP}$&	1.18 (0.99 to 1.43)	&	1.10	&	0.064	\\
			$\alpha_{HsTNT }$&	1.20 (0.98 to 1.56)	&	1.06	&	0.082	\\
			$\alpha_{NT-proBNP }$ &	1.47 (1.21 to 1.79)	&	1.50	&	$<$0.0001	\\	
			$\alpha_{CysC }$&	1.13 (0.73 to 2.60)	&	1.38	&	0.704	\\	
			$\alpha_{NAG }$	&	0.96 (0.63 to 1.34)	&	1.12	&	0.846	\\
			$\alpha_{KIM-1 }$ &1.00 (0.82 to 1.19)	&	1.01	&	0.960	 \\ 
			\hline\noalign{\smallskip}
							
		\end{tabular}}	\caption{Parameter estimates and 95\% credibility intervals under the joint modelling analysis for the Bio-SHiFT data. Hazard ratios are presented per doubling of level, slope or AUC at any point in time.  HR* is the estimate after importance sampling.}  
	\label{tab:3}
	\end{center}
\end{table}
Following adjustment for covariates, the estimated association parameters in Model 1 indicate significant associations between the risk of the composite event and the current underlying values of NT-proBNP and CRP, such that there is a 1.86 fold increase in the risk of the composite event (95\% CI: 1.45 to 2.37), per doubling of NT-probnp level, and a 1.44 fold increase in the risk of the composite event (95\% CI: 1.1 to 1.89), per doubling of CRP level. No significant associations were found for any of HsTNT, CysC, NAG or KIM-1. Similarly, no significant associations were found for the current underlying values of CysC, NAG or KIM-1 in Model 2, and nor were there any significant associations between the risk of the composite event and the slopes of the 6 continuous markers.
Model 3 indicates a significant association for NT-proBNP, with a 1.47 fold increase in the risk of the composite event (95\% CI: 1.21 to 1.79) per doubling of the area under the NT-proBNP profile.
Since the parameter estimates for each of the longitudinal outcomes remained fairly constant across models, to avoid repetition, the estimates and 95\% credibility intervals are presented for one model only (Table~\ref{tab:7}). 

In previous analyses of these same 6 markers, the current underlying value, instantaneous slope and area under the curve of each marker were each assessed independently of one another. Van Boven et al., 2018 found significant associations in all cases for CRP, HsTNT and NT-proBNP, and Brankovic et al., 2018 found significant associations for the current underlying values and slopes of each of CysC, NAG and KIM-1, and the area under the curves for CysC, and NAG. 
Van Boven et al., 2018 provided an additional multivariate analysis for CRP, HsTNT and NT-proBNP, wherein the predicted individual profiles for each marker were separately determined, and functions thereof were simultaneously included in a single extended Cox model as time-varying covariates. Models therefore included either the current underlying values, the instantaneous slopes or the area under the curves for all 3 markers simultaneously. In that analysis, only CRP and NT- proBNP were found to be independently predictive of the composite event, with significant associations for each of the current underlying values and slopes of these markers. In the model for the area under the curves, only NT- proBNP was significant. 


\section{Discussion} 
\label{sec:7}
In this paper, we presented a novel approach for fitting joint models which allows for the inclusion of multivariate longitudinal outcomes with realistic computing times. We demonstrated once again, the bias of the estimated  parameters for the survival process characteristic of the standard two-stage approach, and proposed the use of an importance-sampling corrected two-stage approach, with updated random effects, in its place. Our approach was shown to be successful, producing satisfactory results in a number of simulation scenarios: both survival and longitudinal estimates were unbiased, and computing times were reduced by several orders of magnitude, compared to the full multivariate joint model. We were easily able to incorporate multiple outcomes in the analysis of the Bio-SHiFT data, obtaining very similar results to those previously noted for the CHF-related biomarkers (CRP, HsTNT and NT-proBNP). We did not find any significant associations between any of the renal markers (CysC, NAG and KIM-1) and the risk of the composite event in the multivariate analysis, indicating that their predictive value may not be independent of the CHF-related markers. While the simulations included up to 6 multiple outcomes of varying types, it would be interesting to confirm our results in even more complex settings, (perhaps incorporating competing risks such as those present in the Bio-SHiFT study), and to try determine the limits of the methodology.  A further topic for research would be methods for increasing the speed of computation involved in fitting the multivariate mixed model itself, so as to extend the number of outcomes even further. 
The proposed importance-sampling corrected two-stage estimation approach is implemented in function \texttt{mvJointModelBayes()} in the freely available package \textbf{JMbayes} (version 0.8-0) for the \R \, programming language (freely available from the Comprehensive \R \, Archive Network). An example of how these functions should be used can be found in the appendices.


\section*{Acknowledgements}

The first and last authors acknowledge support by the Netherlands Organization for Scientific Research VIDI grant nr. 016.146.301.\vspace*{-8pt}

\processdelayedfloats
\makeatletter
\efloat@restorefloats
\makeatother

\clearpage

\appendix

\renewcommand\thefigure{\thesection.\arabic{figure}}
\renewcommand\thetable{\thesection.\arabic{table}}
\renewcommand\tablename{Supplementary Table}
\renewcommand\figurename{Supplementary Figure}

	\section{Simulation Study Design}\label{appendix:sim}
	\setcounter{figure}{0}
	\setcounter{table}{0}
\subsection{Scenario I}\label{appendix:simI}
Scenario I simulates 500 patients with a maximum of 15 repeated measurements per patient. We included $K = 2$ continuous longitudinal outcomes and one survival outcome. The $k$ longitudinal outcomes each had form:
\begin{gather*}
y_{ki}(t) = \eta_{ki}(t) +\epsilon_{ki}(t)\\[2pt] = \beta_{k0} + \beta_{k1} \times \textrm{time} + \beta_{k2}\times \textrm{group} + \beta_{k3}\times \textrm{interaction} 
+\,  b_{ki0}+ b_{ki1}\times\textrm{time}+\epsilon_{ki}(t),
\end{gather*}
with $\epsilon_{ki}(t) \sim N(0,\sigma^2_{k})$ and $\textbf{b}_{ki} = (b_{ki0}, b_{ki1})^T$, with $\textbf{b}_i = (\textbf{b}_{1i}^T, \textbf{b}_{2i}^T)^T \sim MVN(\mathbf{0},\textbf{\emph{D}})$. The variance-covariance matrix $\textbf{\emph{D}}$ has general form:
$$\textbf{\emph{D}} = 
\left[\begin{array}{cccc}
\textbf{\emph{D}}_{1} &\multicolumn{2}{c}{\cdots} & D_{else}  \\
\cdots &  \textbf{\emph{D}}_{2} & \multicolumn{2}{r}{\cdots} \\
\multicolumn{2}{l}{\cdots} & \ddots & \multicolumn{1}{r}{\cdots}  \\ 
\multicolumn{3}{l}{\cdots} & \textbf{\emph{D}}_{k} \end{array}\right], \quad
\textbf{\emph{D}}_k =\left[\begin{array}{cccc}
D_{k11} & D_{k12}  \\
D_{k21} & D_{k22}   \end{array}\right]
$$	
and for Scenario I:
$$\textbf{\emph{D}}_{1} = \textbf{\emph{D}}_{2} = \begin{bmatrix}
0.68 &-0.08 \\
-0.08  &0.28  \\
\end{bmatrix}, \textrm{with} \, D_{else} = 0.10$$
Time was simulated from a uniform distribution between 0 and 25. For the survival outcome, adjusting for group allocation, we used:
\begin{gather*}
h_i (t)  = 
h_0(t) \exp \biggl  [\gamma_0 + \gamma_1 \times \textrm{group} + \sum \limits_{k = 1}^K  \alpha_k \eta_{ki}(t) \biggr]\\[2pt]
= 
h_0(t) \exp \biggl  [\gamma_0 + \gamma_1 \times \textrm{group} + \alpha_1 \eta_{1i}(t) + \alpha_2 \eta_{2i}(t)\biggr]. \label{Eq:Surv-RR}
\end{gather*}
The baseline risk was simulated from a Weibull distribution $ h_0(t) = \phi t^{\phi -1}$, with $\phi = 1.65$. For the simulation of the censoring times, an exponential censoring distribution was selected, with mean $\mu = 15$, such that the censoring rate was between $60\%$ and $70\%$. More details are presented in Table~\ref{tab:6}.

\subsection{Scenario II}\label{appendix:simII}
Scenario II is an extension of Scenario I, such that we now have $K = 6$ continuous longitudinal outcomes. We again simulate 500 patients with a maximum of 15 repeated measurements per patient. The $k$ longitudinal outcomes each had form:
\begin{gather*}
y_{ki}(t) =\eta_{ki}(t) +\epsilon_{ki}(t)\\[2pt] = \beta_{k0} + \beta_{k1} \times \textrm{time} + \beta_{k2}\times \textrm{group} + \beta_{k3}\times \textrm{interaction} 
+ \, b_{ki0}+ b_{ki1}\times\textrm{time}+\epsilon_{ki}(t),
\end{gather*}
with $\epsilon_{ki}(t) \sim N(0,\sigma^2_{k})$ and $\textbf{b}_{ki} = (b_{ki0}, b_{ki1})^T$, with $\textbf{b}_i = (\textbf{b}_{1i}^T, \textbf{b}_{2i}^T, \ldots,  \textbf{b}_{6i}^T)^T \sim MVN(\mathbf{0},\textbf{\emph{D}})$, 

\begin{gather*}
\textbf{\emph{D}}_1 = \textbf{\emph{D}}_2 = \begin{bmatrix}
0.97 &0.78 \\
0.07  &0.032  \\
\end{bmatrix}, 
\textbf{\emph{D}}_3 = \textbf{\emph{D}}_4 = \begin{bmatrix}
0.13 &-0.009 \\
-0.009  &0.002  \\
\end{bmatrix},\\
\textbf{\emph{D}}_5 = \textbf{\emph{D}}_6 = \begin{bmatrix}
0.56 &0.05 \\
0.05  &0.03  \\
\end{bmatrix}, \, \textrm{and} \, \textbf{\emph{D}}_{else} = 0.00
\end{gather*}

Time was simulated from a uniform distribution between 0 and 25. For the survival outcome, adjusting for group allocation as in Scenario I, we used:
\begin{eqnarray*}
	\nonumber h_i (t) & = &
	h_0(t) \exp \biggl  [\gamma_1 \times \textrm{group} + \sum \limits_{k = 1}^K  \alpha_k \eta_{ki}(t) \biggr]. 
\end{eqnarray*}
The baseline risk was simulated using B-splines with knots specified apriori. An exponential censoring distribution was used for the simulation of the censoring times, with mean $\mu = 15$, such that the censoring rate was between $60\%$ and $70\%$. Further details are again available in Table~\ref{tab:6}.

\subsection{Scenario III}\label{appendix:simIII}
In Scenario III, we simulate 500 patients with a maximum of 15 repeated measurements per patient, including 3 continuous and 3 binary longitudinal outcomes, such that:
\small
\begin{gather*}
g_k \bigl [ E \{ y_{ki}(t) \mid \bb_{ki} \} \bigr ]= \eta_{ki}(t) \\ = \beta_{k0} + \beta_{k1} \times \textrm{time} + \beta_{k2}\times \textrm{group} + \beta_{k3}\times \textrm{interaction} +
b_{ki0}+ b_{ki1}\times\textrm{time},
\end{gather*}
\normalsize
where $g_k(\cdot)$ denotes the canonical link function appropriate to the response type (identity and logit for the gaussian and binomial outcomes respectively), and $\textbf{b}_i = (\textbf{b}_{1i}^T, \textbf{b}_{2i}^T, \ldots,  \textbf{b}_{6i}^T)^T \sim MVN(\mathbf{0},\textbf{\emph{D}}),$ with

\begin{gather*}
\textbf{\emph{D}}_1 = \textbf{\emph{D}}_2 = \begin{bmatrix}
0.97 &0.78 \\
0.07  &0.032  \\
\end{bmatrix}, 
\textbf{\emph{D}}_3 = \textbf{\emph{D}}_4 = \begin{bmatrix}
0.13 &-0.009 \\
-0.009  &0.002  \\
\end{bmatrix},\\
\textbf{\emph{D}}_5 = \textbf{\emph{D}}_6 = \begin{bmatrix}
0.56 &0.05 \\
0.05  &0.03  \\
\end{bmatrix}, \, \textrm{and} \, \textbf{\emph{D}}_{else} = 0.00
\end{gather*}

For the survival outcome, adjusting for group allocation, we again used:
\small
\begin{gather*}
\nonumber h_i (t)  =
h_0(t) \exp \biggl  [\gamma_1 \times \textrm{group} + \sum \limits_{k = 1}^K  \alpha_k \eta_{ki}(t) \biggr]. 
\end{gather*}
\normalsize
Scenario III maintains the use of the uniform distribution between 0 and 25 for time, and the use of B-splines for the simulation of the baseline hazard. The censoring times were simulated using an exponential censoring distribution as before, with mean $\mu = 15$. Table~\ref{tab:6} provides additional information.

\clearpage

\section{Example R Code}

The below code fits a multivariate joint model for $K = 3$ longitudinal outcomes: $y_1, y_2$ and $y_3$, where $y_1$ is binary and both $y_2$ and $y_3$ are continuous. We fit a linear mixed model for $y_1$ with random intercept and slope (time is $futime$), and use natural cubic splines with two knots, (at $futime = 6$ and $futime = 15$ respectively) in both the fixed and random parts of the models for $y_2$ and $y_3$. The survival submodel adjusts for continuous baseline predictors $x_1$ and $x_2$.

\begin{knitrout}
	\definecolor{shadecolor}{rgb}{0.969, 0.969, 0.969}\color{fgcolor}\begin{kframe}
		\begin{alltt}
			
			\hlcom{# LIBRARIES: JMbayes, splines, survival}
			
			\hlcom{# MULTIVARIATE MIXED MODEL}
			
			MixedModel <- \hlkwd{mvglmer}(list(y1 ~ futime + (futime | id),			
			               y2 ~ \hlkwd{ns}(futime, knots = c(6, 15), Boundary.knots = c(0, 27)) + 
			               (ns(futime, knots = c(6, 15), Boundary.knots = c(0, 27)) | id), 
			               y3 ~ \hlkwd{ns}(futime, knots = c(6, 15), Boundary.knots = c(0, 27)) + 
			               (ns(futime, knots = c(6, 15), Boundary.knots = c(0, 27)) | id)), 
			               data = longit, 
			               families = list(binomial, gaussian, gaussian))
			
			\hlcom{# SURVIVAL SUB MODEL }
			
			SurvFit <- \hlkwd{coxph}(\hlkwd{Surv}(months, pe) ~ x1 + x2, data = surv,	 model = TRUE)
			
			\hlcom{# MULTIVARIATE JOINT MODEL}
			
			JointFitAll <- \hlkwd{mvJointModelBayes}(MixedModel, SurvFit,				timeVar = "futime")
			\hlkwd{summary}(JointFitAll, TRUE) 
			\hlcom{# , TRUE is necessary to obtain the importance-sampling weighted estimates}
			
		\end{alltt}
	\end{kframe}
\end{knitrout}

\clearpage

\section{Tables}
%
\begin{table}[h!]
\begin{center}
	
\scalebox{0.8}{\begin{tabular}{lrrrrrrrrrrrr}
		\hline\noalign{\smallskip}
		 & \multicolumn{4}{c}{Outcome Y1}	& \multicolumn{4}{c}{Outcome Y2} & \multicolumn{4}{c}{Outcome Y3}	\\	 
		\noalign{\smallskip}\hline\noalign{\smallskip}
		Parameter	& True &	Bias	&	RMSE	&	Coverage 
			& True &	Bias	&	RMSE	&	Coverage & 	True & 	Bias	&	RMSE	&	Coverage	\\
		\noalign{\smallskip}\hline\noalign{\smallskip}
		
		Intercept &0.73	&	0.00001	&	0.06	&	0.96	& 0.73&	0.00001	&	0.06	&	0.94 	& 5.75&	0.00000	&	0.02	&	0.96	\\
		Group	& -0.30 &	0.00000	&	0.09	&	0.95	& -0.30&	0.00000	&	0.09	&	0.93 &0.04&	0.00000	&	0.03	&	0.95	\\
		Interaction	& -0.07 &	-0.00001	&	0.02	&	0.92	& -0.07	&	0.00000	&	0.02	&	0.97 & 0.013 &	0.00000	&	0.00	&	0.96	\\
		Time	& 0.24 &	0.00001	&	0.01	&	0.95 & 0.24	&	0.00001	&	0.01	&	0.96 	& 0.07 &	0.00000	&	0.00	&	0.96	\\
		Sigma	& 0.34 &	0.00000	&	0.00	&	0.95	& 0.34	&	0.00001	&	0.00	&	0.96	& 0.19&	0.00000	&	0.00	&	0.96	\\
		
		\noalign{\smallskip}\hline\noalign{\smallskip}
        & \multicolumn{4}{c}{Outcome Y4}	& \multicolumn{4}{c}{Outcome Y5} & \multicolumn{4}{c}{Outcome Y6}	\\	 
        \noalign{\smallskip}\hline\noalign{\smallskip}
        Parameter	& True &	Bias	&	RMSE	&	Coverage 
        & True &	Bias	&	RMSE	&	Coverage & 	True & 	Bias	&	RMSE	&	Coverage	\\
           \noalign{\smallskip}\hline\noalign{\smallskip}		
		
		Intercept	&5.75&	-0.00001	&	0.02	&	0.96	& 10.97&	-0.00001	&	0.06	&	0.95& 10.97&	-0.00001	&	0.07	&	0.94	\\
		Group		& 0.04 &	0.00001	&	0.03	&	0.95	& -0.42&	0.00000	&	0.09	&	0.95&	-0.42& 0.00001	&	0.09	&	0.94	\\
		Interaction		&0.013 &	0.00001	&	0.00	&	0.95	& 0.01 &	0.00000	&	0.02	&	0.95	&0.01 &	0.00001	&	0.02	&	0.95	\\
		Time		&0.07 &	0.00001	&	0.00	&	0.96	& 0.22 &	0.00000	&	0.01	&	0.96	& 0.22 &	0.00001	&	0.01	&	0.95	\\
		Sigma		& 0.19 &	0.00000	&	0.00	&	0.94	& 1.06 &	0.00001	&	0.01	&	0.93	&	1.06 & 0.00000	&	0.01	&	0.94	\\

		\noalign{\smallskip}\hline
	\end{tabular}}
\caption{Simulation results for parameters of the longitudinal submodel (Scenario II)}
\label{tab:4}       
\end{center}
\end{table}

\begin{table}[h!]
\begin{center}
	\scalebox{0.8}{\begin{tabular}{lrrrrrrrrrrrr}
	\hline\noalign{\smallskip}
& \multicolumn{4}{c}{Outcome Y1}	& \multicolumn{4}{c}{Outcome Y2} & \multicolumn{4}{c}{Outcome Y3}	\\	 
\noalign{\smallskip}\hline\noalign{\smallskip}
Parameter	& True &	Bias	&	RMSE	&	Coverage 
& True &	Bias	&	RMSE	&	Coverage & 	True & 	Bias	&	RMSE	&	Coverage	\\
	\noalign{\smallskip}\hline\noalign{\smallskip}

		Intercept & 10.98	&	0.00002	&	0.07	&	0.97	& 10.98&	0.00001	&	0.07	&	0.97	& 10.98&	-0.00001	&	0.05	&	0.93	\\
		Group	& -0.45 &	0.00000	&	0.11	&	0.95& -0.45 &	-0.00001	&	0.11	&	0.96	& -0.45 &	0.00000	&	0.07	&	0.94	\\
		Interaction	& 0.05 &	-0.00001	&	0.02	&	0.94		&  0.05 &	-0.00001	&	0.02	&	0.95	&  0.05 &	0.00000	&	0.01	&	0.93	\\
		Time	& 0.21 &	0.00002	&	0.01	&	0.95		& 0.21 &	0.00001	&	0.01	&	0.96	& 0.21 &	0.00001	&	0.01	&	0.93	\\
		Sigma	& 1.10 &	-0.00001	&	0.01	&	0.95	&  1.10 &	0.00002	&	0.01	&	0.96	&  1.10 &	0.00000	&	0.01	&	0.94	\\
		\noalign{\smallskip}\hline\noalign{\smallskip}

& \multicolumn{4}{c}{Outcome Y4}	& \multicolumn{4}{c}{Outcome Y5} & \multicolumn{4}{c}{Outcome Y6}	\\	 
\noalign{\smallskip}\hline\noalign{\smallskip}
Parameter	& True &	Bias	&	RMSE	&	Coverage 
& True &	Bias	&	RMSE	&	Coverage & 	True & 	Bias	&	RMSE	&	Coverage	\\
\noalign{\smallskip}\hline\noalign{\smallskip}		
		
		Intercept	& 1.11&	-0.00002	&	0.10	&	0.94& 1.11&		0.00000	&	0.12	&	0.95	& 1.11&		-0.00001	&	0.12	&	0.95	\\
		Group	&	-1.09 &0.00001	&	0.13	&	0.95	&-1.09 &	0.00000	&	0.15	&	0.97		&-1.09 &	-0.00001	&	0.15	&	0.96	\\
		Interaction	&0.01 &	0.00002	&	0.02	&	0.93	&	0.01 &	0.00000	&	0.03	&	0.96	&	0.01 &	0.00001	&	0.03	&	0.96	\\
		Time	& 0.14 &	0.00001	&	0.02	&	0.93	& 0.14 &	-0.00002	&	0.03	&	0.94	& 0.14 &	0.00002	&	0.03	&	0.93	\\
		
		\noalign{\smallskip}\hline
	\end{tabular}}
\caption{Simulation results for parameters of the longitudinal submodel (Scenario III)}
\label{tab:5}       
\end{center}
\end{table}

\clearpage
\begin{table}[h!]
\begin{center}
	\begin{tabular}{llllll}
		\noalign{\smallskip}\hline
		Scenario & Y & $\alpha_k$ & $\bfbeta_k = (\beta_{k0}, \beta_{k1}, \beta_{k2}, \beta_{k3})$ & $\sigma_k$ & $\bfgamma_k = (\gamma_{k0}, \gamma_{k1})$ \\ 
		
		\noalign{\smallskip}\hline\noalign{\smallskip}
		
		\multirow{2}{*}{I} & 1 & 0.64 & \multirow{2}{*} {(2.13, 0.24, -0.25, -0.05)} & \multirow{2}{*}{0.6} & \multirow{2}{*}{(-5.8, 0.48)}  \\ 
		& 2 & -0.64 &  & &   \\ 
		\noalign{\smallskip}\hline\noalign{\smallskip}
		
		\multirow{6}{*}{II} & 1 &  1.09 &\multirow{2}{*} {(0.73, 0.24, -0.30, -0.07)} & \multirow{2}{*}{0.34} & \multirow{2}{*}{(na, -0.2)}   \\ 
		& 2  & -1.09 &  & & \\
		\noalign{\smallskip}\cline{2 - 6} \noalign{\smallskip}
		& 3  & -0.92 &\multirow{2}{*} {(5.75, 0.07, 0.04, 0.013)} & \multirow{2}{*}{0.19} & \multirow{2}{*}{(na, -0.2)}   \\ 
		& 4  & 0.92 &  & &  \\ 
		\noalign{\smallskip}\cline{2 - 6} \noalign{\smallskip}
		& 5  & 0.43 &\multirow{2}{*} {(10.97, 0.22, -0.42, 0.01)} & \multirow{2}{*}{1.06} & \multirow{2}{*}{(na, -0.2)}   \\ 
		& 6  & -0.43 &  & &   \\  
		\noalign{\smallskip}\hline\noalign{\smallskip}
		\multirow{6}{*}{III} & 1 &  0.17 &\multirow{3}{*} {(10.98, 0.21, -0.45, 0.05)} & \multirow{3}{*}{1.1} & \multirow{3}{*}{(na, -0.47)}   \\ 
		& 2  & -0.17 &  & & \\
		& 3  & 0.17 &  & &   \\ 
		\noalign{\smallskip}\cline{2 - 6} \noalign{\smallskip}
		& 4  & -0.17  & \multirow{3}{*} {(1.11, 0.14, -1.09, 0.01)} & \multirow{3}{*} {na} & \multirow{3}{*}{(na, -0.47)} \\
		& 5  & 0.47 &  &  & \\  
		& 6  & -0.47 &  &  & \\  
		\noalign{\smallskip}\hline
	\end{tabular} 
\caption{Simulation Scenarios}
\label{tab:6}       
\end{center}
\end{table}

\clearpage

\begin{table}[h!]
\begin{center}
\scalebox{0.7}{\begin{tabular}{lrrrrrrrrr}
		\noalign{\smallskip}\hline
		\multicolumn{4}{c}{Longitudinal Process} \\
		\noalign{\smallskip}\hline\noalign{\smallskip}
		
		&	\multicolumn{3}{c}{CystatinC}	&\multicolumn{3}{c}{NAG} 		&\multicolumn{3}{c}{KIM-1}		\\ 
		&	Est. (95\% CI) & Est.*	&	p-value	&	Est. (95\% CI) & Est.*	&	p-value	&	Est. (95\% CI) & Est.*	&	p-value		\\ 
		\noalign{\smallskip}\cline{2-10}\noalign{\smallskip} 
		
		Intercept &	-0.36 (-0.42 to -0.31)	&	-0.38	&	$<$0.0001&	2.46 (2.34 to 2.58)	&	2.42	&	$<$0.0001 &	8.93 (8.78 to 9.07) 	&	8.99	&	$<$0.0001	 \\
		Time (years since baseline)&	-0.04 (-0.07 to -0.01)	&	-0.05	&	0.004&	-0.06 (-0.13 to 0.00)	&	-0.07	&	0.058 &	0.01 (-0.05 to 0.08)	&	0.02	&	0.69	\\
		$\sigma$&	0.43 (0.42 to 0.45)	&	0.43	&	$<$0.0001	&	0.93 (0.90 to 0.97)	&	0.93	&	$<$0.0001&	0.85 (0.82 to 0.88)	&	0.81	&	$<$0.0001	 \\ 
		\noalign{\smallskip}\noalign{\smallskip} 
		
		&	\multicolumn{3}{c}{CRP}			&	\multicolumn{3}{c}{HsTNT} 	&	\multicolumn{3}{c}{NTS-proBNP}	\\ 
		&	Est. (95\% CI) & Est.*	&	p-value 	&	Est. (95\% CI) & Est.*	&	p-value	&	Est. (95\% CI) & Est.*	&	p-value\\ 
		\noalign{\smallskip}\cline{2-10}\noalign{\smallskip}  
		
		Intercept	&	1.06 (0.87 to 1.26)	&	1.05	&	$<$0.0001&	4.17 (4.02 to 4.32)	&	4.17	&	$<$0.0001 	&	6.78 (6.54 to 7.01)	&	6.59	&$<$0.0001	 \\
		$B_n(Time, \lambda_1)$ 	&	0.96 (0.65 to 1.26)	&	1.18	&	$<$0.0001&	0.17 (0.05 to 0.29)	&	0.26	&	0.01 	&	0.00 (-0.17 to 0.18)	&	0.12	&	1.00	\\
		$B_n(Time, \lambda_2)$ 	&	0.50 (0.31 to 0.69)	&	0.35	&	$<$0.0001 &	0.17 (0.10 to 0.26)	&	0.17	&	$<$0.0001 &	-0.04 (-0.30 to 0.23)	&	0.05	&	0.75	\\
		$B_n(Time, \lambda_3)$ & na & na & na & na & na & na &	0.17 (0.03 to 0.34)	&	0.23	&	0.01	\\
		$\sigma$	&	1.00 (0.96 to 1.04)	&	0.99	&	$<$0.0001&	0.28 (0.27 to 0.29)	&	0.29	&	$<$0.0001 &	0.50 (0.48 to 0.52)	&	0.50	&	$<$0.0001	\\
        \noalign{\smallskip}\hline
		
	\end{tabular}}
	\caption{Parameter estimates and 95\% credibility intervals under the joint modelling analysis for the Bio-SHiFT data (Model 1).  Est.* are the estimates after importance sampling.}  
\label{tab:7}
\end{center}
	
\end{table}

\clearpage

\section{Figures}

\begin{figure}[h!]
	\begin{center}
		\includegraphics[width = 0.75\textwidth]{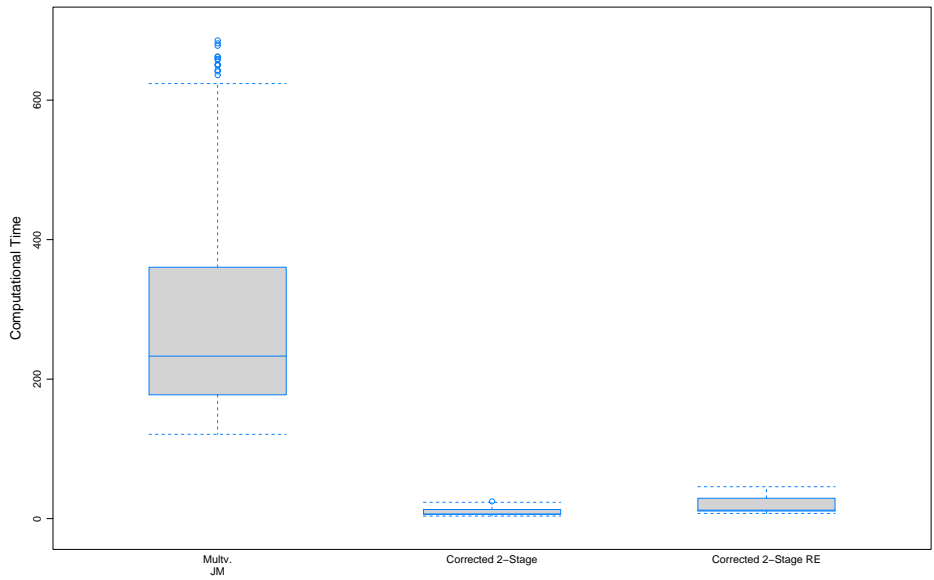}
	\end{center}
	\caption{Simulation results from 500 datasets comparing the importance-sampling-corrected two-stage approach with and without updated random effects with the full multivariate joint model. The boxplots show the mean computational time per approach (in minutes).} \label{fig:4}
\end{figure}

\clearpage

\begin{figure}[h!]
	\begin{center}
		\includegraphics[width = 0.75\textwidth]{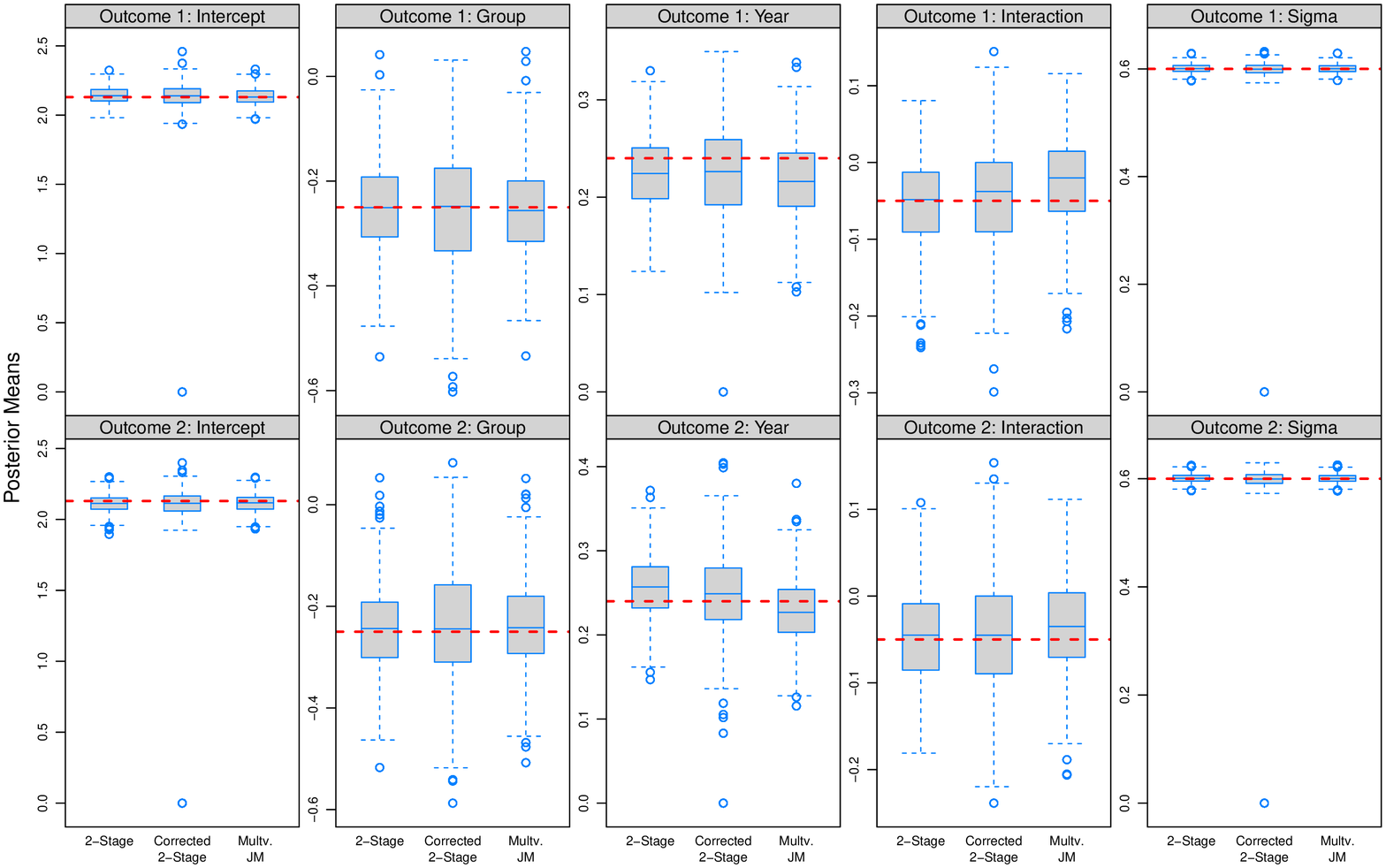}
	\end{center}
	\caption{Simulation results from 500 datasets comparing the simple two-stage approach and the importance-sampling-corrected two-stage approach with the full multivariate joint model. The  panels show the posterior means from the 500 datasets for the coefficients from the two longitudinal outcomes in Scenario I. The dashed horizontal line indicates the true value of the coefficients.} \label{fig:5}
\end{figure}

\clearpage	

\begin{figure}[h!]
	\begin{center}
		\includegraphics[width = 0.75\textwidth]{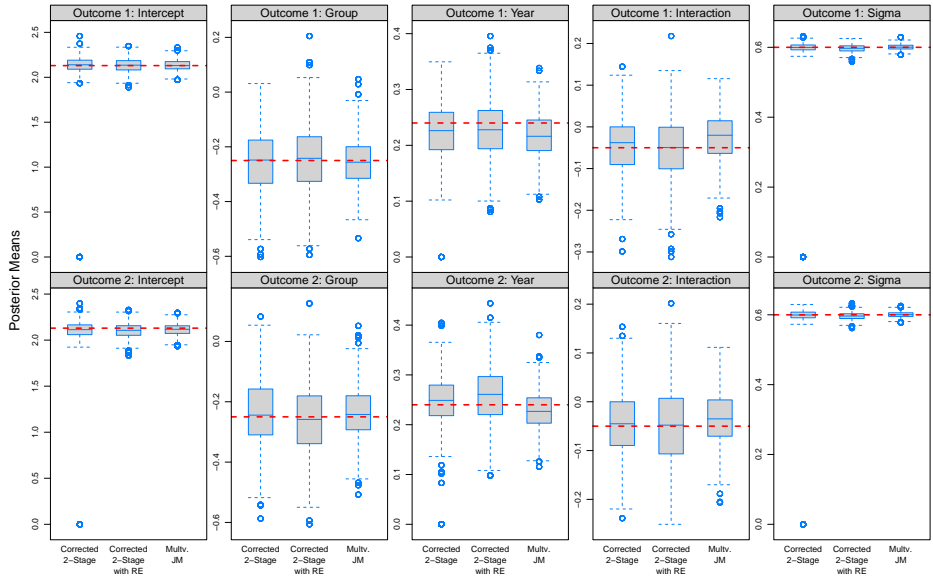}
	\end{center}
	\caption{Simulation results from 500 datasets comparing the importance-sampling-corrected two-stage approach with and without updated random effects with the full multivariate joint model. The  panels show the posterior means from the 500 datasets for the coefficients from the two longitudinal outcomes in Scenario I. The dashed horizontal line indicates the true value of the coefficients.} \label{fig:6}
\end{figure}

\clearpage

\begin{figure}[h!]
	\begin{center}
		\includegraphics[width = 0.75\textwidth]{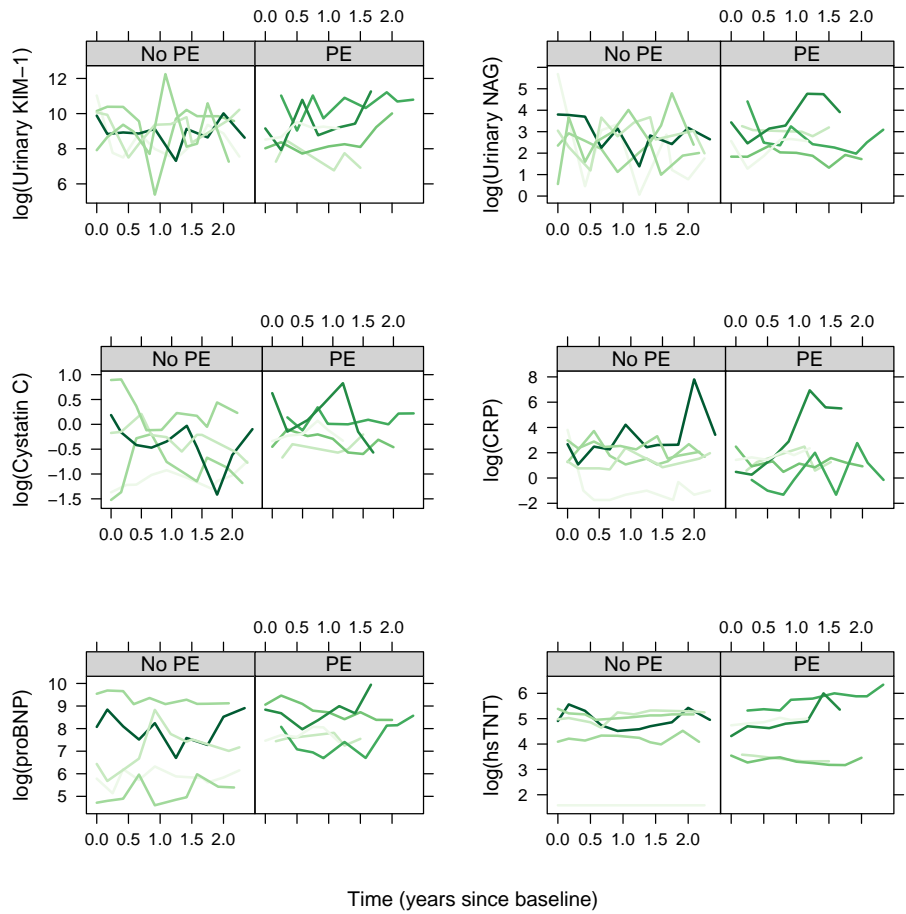}
	\end{center}
	\caption{Longitudinal profiles of continuous biomarkers (log base 2) for a randomly selected subset of individuals from the Bio-SHiFT cohort study that did/did not experience the primary event of interest.} \label{fig:7}
\end{figure}

\clearpage

\renewcommand\refname{Bibliography}

\bibliographystyle{unsrtnat} 
\bibliography{MultJM} 

\begin{thebibliography}{27}
\providecommand{\natexlab}[1]{#1}
\providecommand{\url}[1]{\texttt{#1}}
\expandafter\ifx\csname urlstyle\endcsname\relax
  \providecommand{\doi}[1]{doi: #1}\else
  \providecommand{\doi}{doi: \begingroup \urlstyle{rm}\Url}\fi

\bibitem[Faucett and Thomas(1996)]{faucett.thomas:96}
C.~Faucett and D.~Thomas.
\newblock Simultaneously modelling censored survival data and repeatedly
  measured covariates: {A} {G}ibbs sampling approach.
\newblock \emph{Statistics in Medicine}, 15:\penalty0 1663--1685, 1996.

\bibitem[Wulfsohn and Tsiatis(1997)]{wulfsohn.tsiatis:97}
M.~Wulfsohn and A.~Tsiatis.
\newblock A joint model for survival and longitudinal data measured with error.
\newblock \emph{Biometrics}, 53:\penalty0 330--339, 1997.

\bibitem[Brown et~al.(2005{\natexlab{a}})Brown, Ibrahim, and
  DeGruttola]{brown.et.al:05}
E.~R. Brown, J.~G. Ibrahim, and V.~DeGruttola.
\newblock A flexible {B}-spline model for multiple longitudinal biomarkers and
  survival.
\newblock \emph{Biometrics}, 61:\penalty0 64--73, 2005{\natexlab{a}}.

\bibitem[Elashoff et~al.(2008)Elashoff, Li, and Li]{elashoff.et.al:08}
R.~Elashoff, G.~Li, and N.~Li.
\newblock A joint model for longitudinal measurements and survival data in the
  presence of multiple failure types.
\newblock \emph{Biometrics}, 64:\penalty0 762--771, 2008.

\bibitem[Andrinopoulou et~al.(2014)Andrinopoulou, Rizopoulos, Takkenberg, and
  Lesaffre]{andrinopoulou.et.al:14}
E.~R. Andrinopoulou, D.~Rizopoulos, J.~Takkenberg, and E.~Lesaffre.
\newblock Joint modeling of two longitudinal outcomes and competing risk data.
\newblock \emph{Statistics in Medicine}, 33:\penalty0 3167--3178, 2014.

\bibitem[Ferrer et~al.(2016)Ferrer, Rondeau, Dignam, Pickles, Jacqmin-Gadda,
  and Proust-Lima]{ferrer.et.al:16}
L.~Ferrer, V.~Rondeau, J.~Dignam, T.~Pickles, H.~Jacqmin-Gadda, and
  C.~Proust-Lima.
\newblock Joint modelling of longitudinal and multi-state processes:
  application to clinical progressions in prostate cancer.
\newblock \emph{Statistics in Medicine}, 35:\penalty0 3933--3948, 2016.

\bibitem[Proust-Lima and Taylor(2009)]{proust-lima.taylor:09}
C.~Proust-Lima and J.~M.~G. Taylor.
\newblock Development and validation of a dynamic prognostic tool for prostate
  cancer recurrence using repeated measures of posttreatment {PSA}: {A} joint
  modeling approach.
\newblock \emph{Biostatistics}, 10:\penalty0 535--549, 2009.

\bibitem[Rizopoulos(2011)]{rizopoulos:11}
D.~Rizopoulos.
\newblock Dynamic predictions and prospective accuracy in joint models for
  longitudinal and time-to-event data.
\newblock \emph{Biometrics}, 67:\penalty0 819--829, 2011.

\bibitem[Rizopoulos et~al.(2014)Rizopoulos, Hatfield, Carlin, and
  Takkenberg]{rizopoulos.et.al:14}
D.~Rizopoulos, L.~Hatfield, B.~Carlin, and J.~Takkenberg.
\newblock Combining dynamic predictions from joint models for longitudinal and
  time-to-event data using {B}ayesian model averaging.
\newblock \emph{JASA}, 109:\penalty0 1385--1397, 2014.

\bibitem[Andrinopoulou and Rizopoulos(2016)]{andrinopoulou.rizopoulos:16}
{E. ~R.} Andrinopoulou and D.~Rizopoulos.
\newblock Bayesian shrinkage approach for a joint model of longitudinal and
  survival outcomes assuming different association structures.
\newblock \emph{Statistics in Medicine}, 35:\penalty0 4813--4823, 2016.

\bibitem[Rizopoulos et~al.(2017)Rizopoulos, Molenberghs, and
  Lesaffre]{rizopoulos.et.al:17}
D.~Rizopoulos, G.~Molenberghs, and E.~M. E.~H. Lesaffre.
\newblock Dynamic predictions with time-dependent covariates in survival
  analysis using joint modeling and landmarking.
\newblock \emph{Biometrical Journal}, 59:\penalty0 1261--1276, 2017.

\bibitem[Andrinopoulou et~al.(2018)Andrinopoulou, Eilers, Takkenberg, and
  Rizopoulos]{andrinopoulou.et.al:18}
E.~R. Andrinopoulou, P.~H.~C. Eilers, J.~J.~M. Takkenberg, and D.~Rizopoulos.
\newblock Improved dynamic predictions from joint models of longitudinal and
  survival data with time-varying effects using p-splines.
\newblock \emph{Biometrics}, 00:\penalty0 00--00, 2018.
\newblock \doi{10.1111/biom.12814}.

\bibitem[Rizopoulos and Ghosh(2011{\natexlab{a}})]{RizGhosh2011}
D.~Rizopoulos and P.~Ghosh.
\newblock A {Bayesian} semiparametric multivariate joint model for multiple
  longitudinal outcomes and a time-to-event.
\newblock \emph{Statistics in Medicine}, 30:\penalty0 1366--1380,
  2011{\natexlab{a}}.

\bibitem[Chi and Ibrahim(2006)]{ChiIb2006}
Yueh-Yun Chi and Joseph~G Ibrahim.
\newblock Joint models for multivariate longitudinal and multivariate survival
  data.
\newblock \emph{Biometrics}, 62\penalty0 (2):\penalty0 432--445, 2006.

\bibitem[Brown et~al.(2005{\natexlab{b}})Brown, Ibrahim, and
  DeGruttola]{BrownIbDeG2005}
Elizabeth~R Brown, Joseph~G Ibrahim, and Victor DeGruttola.
\newblock A flexible b-spline model for multiple longitudinal biomarkers and
  survival.
\newblock \emph{Biometrics}, 61\penalty0 (1):\penalty0 64--73,
  2005{\natexlab{b}}.

\bibitem[Lin et~al.(2002)Lin, McCulloch, and Mayne]{Lin2002}
Haiqun Lin, Charles~E McCulloch, and Susan~T Mayne.
\newblock Maximum likelihood estimation in the joint analysis of time-to-event
  and multiple longitudinal variables.
\newblock \emph{Statistics in Medicine}, 21\penalty0 (16):\penalty0 2369--2382,
  2002.

\bibitem[van Boven et~al.(2018)van Boven, Battes, Akkerhuis, Rizopoulos,
  Caliskan, Anroedh, et~al.]{bioshift}
N.~van Boven, L.~C. Battes, K.~M. Akkerhuis, D.~Rizopoulos, K.~Caliskan, S.~S.
  Anroedh, et~al.
\newblock Toward personalized risk assessment in patients with chronic heart
  failure: Detailed temporal patterns of nt-probnp, troponin t, and crp in the
  bio-shift study.
\newblock \emph{American Heart Journal}, 196:\penalty0 36--48, 2018.

\bibitem[Tsiatis and Davidian(2004)]{tsiatis.davidian:04}
A.~A. Tsiatis and M.~Davidian.
\newblock Joint modeling of longitudinal and time-to-event data: An overview.
\newblock \emph{Statistica Sinica}, 14:\penalty0 809--834, 2004.

\bibitem[Rizopoulos(2012)]{rizopoulos:12}
D.~Rizopoulos.
\newblock \emph{Joint Models for Longitudinal and Time-to-Event Data, with
  Applications in R}.
\newblock Chapman \& Hall/CRC, Boca Raton, 2012.

\bibitem[Ye et~al.(2008)Ye, Lin, and Taylor]{ye.et.al:08b}
W.~Ye, X.~Lin, and J.~Taylor.
\newblock Semiparametric modeling of longitudinal measurements and
  time-to-event data -- a two stage regression calibration approach.
\newblock \emph{Biometrics}, 64:\penalty0 1238--1246, 2008.

\bibitem[Press et~al.(2007)Press, Teukolsky, Vetterling, and
  Flannery]{press.et.al:07}
W.~Press, S.~Teukolsky, W.~Vetterling, and B.~Flannery.
\newblock \emph{Numerical Recipes: The Art of Scientific Computing}.
\newblock Cambridge University Press, New York, 3rd edition, 2007.

\bibitem[Brown(2009)]{brown:09}
E.~R. Brown.
\newblock Assessing the association between trends in a biomarker and risk of
  event with an application in pediatric {HIV/AIDS}.
\newblock \emph{The Annals of Applied Statistics}, 3:\penalty0 1163--1182,
  2009.

\bibitem[Rizopoulos and Ghosh(2011{\natexlab{b}})]{rizopoulos.ghosh:11}
D.~Rizopoulos and P.~Ghosh.
\newblock A {Bayesian} semiparametric multivariate joint model for multiple
  longitudinal outcomes and a time-to-event.
\newblock \emph{Statistics in Medicine}, 30:\penalty0 1366--1380,
  2011{\natexlab{b}}.

\bibitem[Lewandowski et~al.(2009)Lewandowski, Kurowicka, and
  Joe]{lewandowski.et.al:09}
D.~Lewandowski, D.~Kurowicka, and H.~Joe.
\newblock Generating random correlation matrices based on vines and extended
  onion method.
\newblock \emph{Journal of Multivariate Analysis}, 100:\penalty0 1989--2001,
  2009.

\bibitem[Lang and Brezger(2004)]{lang.brezger:04}
S.~Lang and A.~Brezger.
\newblock Bayesian {P}-splines.
\newblock \emph{Journal of Computational and Graphical Statistics},
  13:\penalty0 183--212, 2004.

\bibitem[Jullion and Lambert(2007)]{jullion.lambert:07}
A.~Jullion and P.~Lambert.
\newblock Robust specification of the roughness penalty prior distribution in
  spatially adaptive bayesian p-splines models.
\newblock \emph{Computational Statistics and Data Analysis}, 51:\penalty0
  2542--2558, 2007.

\bibitem[Brankovic et~al.(2018)Brankovic, Akkerhuis, van Boven, Anroedh,
  Constantinescu, Caliskan, et~al.]{kidney}
M.~Brankovic, M.~Akkerhuis, N.~van Boven, S.~Anroedh, A.~Constantinescu,
  K.~Caliskan, et~al.
\newblock Patient-specific evolution of renal function dynamically predicts
  clinical outcome in chronic heart failure: the bio- shift study.
\newblock \emph{Kidney International}, 4:\penalty0 952--960, 2018.
\newblock URL \url{https://www.ncbi.nlm.nih.gov/pubmed/29191357}.

\end{thebibliography}

\end{document}